\documentclass[3p]{elsarticle}

\usepackage{lineno,hyperref}
\modulolinenumbers[200]

\usepackage{bm}
\usepackage[cmex10]{amsmath}
\newcommand{\refeq}[1]{Eq.\,(\ref{#1})} 
\newcommand\TE{\mathit{TE}} 
\newcommand\MHz{\mathit{MHz}}
\newcommand\ADC{\mathit{ADC}} 
\DeclareMathOperator{\pinv}{pinv}
\usepackage{amssymb}
\usepackage{url}
\journal{ArXiv}

\biboptions{authoryear}

\begin{document}

\begin{frontmatter}

\title{Diffusion MRI microstructure models with \emph{in vivo} human brain Connectom data: results from a multi-group comparison}

 \author[ucl1,nyu,ucl2]{Uran Ferizi\corref{cor1}}
 \ead{uran.ferizi@med.nyu.edu}
 \author[crl]{Benoit Scherrer\corref{cor1}}
 \author[ucl2,phl]{Torben Schneider\corref{cor1}}

 \address[ucl1]{Centre for Medical Image Computing, University College London, UK}
  \address[nyu]{Department of Radiology, New York University School of Medicine, USA}
 \address[ucl2]{Department of Neuroinflammation, Institute of Neurology, University College London, UK}
 \address[crl]{Computational Radiology Laboratory, Boston Children's Hosp., Harvard University, USA}
\address[phl]{Philips Healthcare, Guildford, Surrey, UK}

 \cortext[cor1]{Joint first co-authors}

\author[gothenburg]{Mohammad Alipoor}
\address[gothenburg]{Chalmers University of Technology, Gothenburg, Sweden}

\author[guanajuato]{Odin Eufracio}
 \address[guanajuato]{Centro de Investigacion en Matematicas AC, Guanajuato, Mexico}

\author[antipolis]{Rutger H.J. Fick}
 \author[antipolis]{Rachid Deriche}
\address[antipolis]{Athena Project-Team, INRIA Sophia Antipolis - M\'editerran\'ee, France}

\author[lund]{Markus Nilsson}
 \address[lund]{Lund University Bioimaging Center, Lund University, Sweden}

\author[guanajuato]{Ana K. Loya-Olivas}
\author[guanajuato]{Mariano Rivera}

\author[erasmus]{Dirk H.J. Poot}
 \address[erasmus]{Erasmus Medical Center and Delft University of Technology, the Netherlands}

\author[guanajuato]{Alonso Ramirez-Manzanares}
\author[guanajuato]{Jose L. Marroquin}

\author[Washington]{Ariel Rokem}
\address[Washington,stanford]{eScience Institute, University of Washington, USA}
\author[stanford]{Christian P\"{o}tter}
\author[stanford]{Robert F. Dougherty}
\address[stanford]{Center for Cognitive and Neurobiological Imaging, Stanford University, USA}

\author[cleveland]{Ken Sakaie}
 \address[cleveland]{Imaging Institute, The Cleveland Clinic, Cleveland, USA}

\author[ucl2]{Claudia Wheeler-Kingshott}
 \author[crl]{Simon K. Warfield} 
 \author[mit]{Thomas Witzel}
 \author[mit]{Lawrence L. Wald}
 \author[nyu]{Jos\'e G. Raya} 
 \author[ucl1]{Daniel C. Alexander}

 \address[mit]{A.A. Martinos Center for Biomedical Imaging, MGH, Harvard University, USA}


%
%

\begin{abstract}
A large number of mathematical models have been proposed to describe the measured signal in diffusion-weighted (DW) magnetic resonance imaging (MRI). However, model comparison to date focuses only on specific subclasses, and little or no information is available in the literature on how performance varies among the different types of models.
To address this deficiency we organized the ``White Matter Modeling Challenge'' during the International Symposium on Biomedical Imaging (ISBI) 2015 conference.  
This competition aimed to compare a range of different kinds of model in their ability to explain measured \textit{in vivo} DW human brain data. We focus specifically on the challenge of explaining a large range of measurable data. We used the Connectome scanner at the Massachusetts General Hospital, using gradients strengths of up to $300$ mT/m and a broad set of diffusion times.
We focused on assessing the DW signal prediction in two regions: the genu in the corpus callosum, where the fibres are relatively straight and parallel, and the fornix, where the configuration of fibres is more complex.  The challenge participants had access to three-quarters of the dataset, and their models  were ranked on their ability to predict the remaining unseen quarter of the data. The challenge provided a unique opportunity for quantitative comparison of a diverse set of methods from multiple groups worldwide.
The comparison of the challenge entries reveals important trends and conclusions that influence the next generation of diffusion-based quantitative MRI techniques. 
The first is that signal models do not necessarily outperform tissue models; in fact tissue models on average rank highest of those tested. The second is that assuming a non-Gaussian (rather than a purely Gaussian) noise model provides little benefit. The third is that preprocessing the training data (here, omitting signal outliers) and using signal predicting strategies, such as bootstrapping or cross-validation, could benefit the model fitting.
The analysis in this study provides a benchmark for other models and the data remains available to build up a more complete comparison over future years.

\end{abstract}


\end{frontmatter}

\section{Introduction}
Diffusion-weighted (DW) magnetic resonance imaging (MRI) can provide unique insights into the microstructure of living tissue and is increasingly used to study the microanatomy and development of normal functioning tissue as well as its pathology. Mathematical models for analysis and interpretation have been crucial for the clinical adoption of DW-MRI. 
Even though diffusion tensor imaging (DTI) \citep{basser}, which is based on a simple model of the DW-MRI signal, has shown promise in clinical applications \citep{Assaf2008}, e.g. Alzheimer's disease \citep{rose2000loss}, Multiple Sclerosis \citep{werring2000pathogenesis} or brain tumors \citep{price2006improved}, a much wider variety of DW-MRI models has been proposed to extract more information from the DW signal. 

Models generally fall between two extremes of “models of the tissue” and “models of the signal”.Models of the tissue \citep{behrens2003characterization,behrens2007probabilistic,assaf2005composite,activeax,dyrby,sotiropoulos,zhang_noddi,scherrer_mrm2015,kaden2016multi} explicitly describe the underlying tissue microstructure in each voxel with a multi-compartment approach \citep{stanisz,panagiotaki,nilsson2013role}. 
Models of the signal focus on describing the DW signal attenuation without explicitly describing the underlying tissue composition that gives rise to the signal \citep{liu2003generalized,Tuch2004,jensen2005diffusional,jensen,Descoteaux2007,assemlal,yablonskiy2010theoretical,kiselev2011cumulant,shore,ozarslan2013mean}.
Other approaches fall between these two classes and include some features of the tissue, such as the distribution of fibre orientations, but often describe the signal from individual fibres without modelling the fibre composition explicitly \citep{mitra1992diffusion,Tournier2004,alexander2005maximum,anderson2005measurement,jian2007multi,jian2007novel,sakaie2007objective,dell2007model,jbabdi,rathi2014multi,kaden2015quantitative}.   

Despite this explosion of DW-MRI models, a broad comparison on a common dataset and within a common evaluation framework is lacking, so little is understood about which models are more plausible representations or explanations of the signal.  
\citet{panagiotaki} established a taxonomy of diffusion compartment models and compared 47 of them using data from the fixed corpus callosum of a rat acquired on a pre-clinical system. Later, \citet{ferizi_mrm} performed a similar experiment using data from a live human subject, while \citet{ferizi_miccai,ferizi_nimg} explored a different class of models that aim to capture fiber dispersion. \citet{Rokem2015} compared two classes of models using cross-validation and test-retest accuracy.
All these studies \citep{panagiotaki,ferizi2014compartment,Rokem2015} aim to evaluate variations with specific classes of models with all other variables of the parameter estimation pipeline (i.e. noise model, fitting routine, etc.) fixed.  While this provides fundamental insight into which compartments are important in compartment models, questions remain about the broader landscape of models; in particular, which classes of models explain the signal best and how strongly performance depends on the choice of parameter-estimation procedure.

Publicly organized challenges provide a unique opportunity to bring a research community together to gain a quantitative and unbiased comparison of a diverse set of methods applicable to a particular data-processing task. Such publicly organized challenges have helped to establish a common ground for the evaluation of competing methods in a variety of imaging-related tasks, e.g. registration of MRI brain images \citep{klein2009evaluation}, diagnostic group classification for dementia using structural MRI \citep{bron2015standardized}, tissue segmentation on brain \citep{Mendrik2015} and prostate tissue \citep{Litjens2014}, on CT images for thoracic tissue \citep{murphy2011evaluation}, carotid tissue \citep{Hameeteman2011}, and breathing airways tissue \citep{Lo2012}, fetal ultrasound images \citep{Rueda2014}, or particle tracking \citep{Chenouard2014} have been organized. 
In DW-MRI, public challenges have focused on recovering synthetic intra-voxel fibre configurations \citep{daducci2014quantitative} or evaluating tractography techniques \citep{fillard2011quantitative,Pujol2015} and have been very successful at driving research and translation forward. Another interesting comparison of reconstruction methods using DW-MRI data was based on signal acquired from a physical phantom \citep{ningsparse}.

Here we report on such a community-wide challenge to model the variation of DW-MRI signals at the voxel level in the \emph{in vivo} human brain.
Modelling the diffusion signal is a key step in realising practical and reliable quantitative imaging techniques based on diffusion MRI. The challenge in the area is to extract the salient features from the diffusion signal and relate them to the principal features of the underlying tissue (e.g., in the case of brain white matter, the fibre orientation, axonal packing and axonal size). Three distinct questions arise: 
i) given the richest possible dataset that samples the space of achievable measurements as widely as possible, which mathematical model can capture best the intrinsic variation of the acquired signal; 
ii) which tissue features can be derived from the model;
iii) what subset of those features can we estimate from limited acquisition time on a standard clinical scanner and what dataset best supports such estimates?
The intuition gained from (i) is generalisable over a wide range of applications, while (ii) and (iii) are highly dependent on the MRI study design and the available hardware. Therefore, our challenge focuses on question (i), as an understanding of (i) is necessary to inform (ii) and (iii). 
To that end, we acquire the richest possible dataset using the most powerful hardware available and the most motivated subject available (UF). Specifically, we use the Connectome scanner \citep{connectom}, which is unique among human scanners in having 300 mT/m gradients, rather than 40 mT/m typical of state-of-the-art human scanners. Preclinical work by \citet{dyrby} highlights the benefits of such strong gradients and the first results from the Connectome scanner \citep{mcnab,duval,ferizi_nimg,huang} are now starting to verify those findings.

The uniquely rich dataset from \citet{ferizi_nimg}, acquired on the Connectome system, samples around five thousand points in the space of possible measurements from a standard Stejskal-Tanner DW-MRI sequence. Each DWI has a unique combination of gradient strength, diffusion time, pulse width and echo time; i.e. they vary all the key parameters of the sequence to highlight sensitivity to as many underlying tissue properties as possible. This offers a unique opportunity for the comparison of the many different types of models within a common framework, over a very wide range of the measurement space. 
Using this rich dataset we organized the White Matter Modeling challenge, held during the 2015 International Symposium on Biomedical Imaging (ISBI) in New York. The goal of the challenge was to evaluate and compare the models in two different tissue configurations that are common in the brain: 1) a white matter region of interest where fibers are relatively straight and parallel, specifically, the genu of the corpus callosum; and 2) a region in which the fiber configuration is more complex, specifically, the fornix.
Challenge participants had access to three-quarters of each whole dataset; the participating models  were evaluated on how well they predicted the  remaining `unseen' part of the data. 
This kind of model comparison, based on prediction error, is a common and crucial part of the development of any statistical model-based estimation applications. Books such as \citet{burnhamanderson} explain how and why such comparisons should be performed to reject models that are theoretically plausible but that the data do not support. As announced before the challenge, the final ranking was based exclusively on the performance on the genu data. In this paper, however, we include results from both the genu and the fornix.

The challenge entries include a wide range of different kinds of model and estimation procedure. In contrast to earlier model comparisons \citep{panagiotaki,ferizi2014compartment,Rokem2015}, the results provide new insight into which broad classes of model explain the signal best and what features of the estimation procedure are important. Although the sampling of the set of possible techniques is necessarily sparse, as any model could in theory combine with any estimation procedure and each has many variables, the results provide some surprising and key insights into the benefits of different approaches. This information is very timely, as recent model-based diffusion MRI techniques, such as NODDI \citep{zhang_noddi}, SMT \citep{kaden2015quantitative,kaden2016multi}, DIAMOND \citep{scherrer_mrm2015}, DKI \citep{fieremans2011white} and LEMONADE \citep{lemonade}, are starting to become widely adopted in clinical studies and trials. Despite their success, intense debate continues in the field about applicability of different models and fitting routines \citep{jelescu2015one,jelescu2016degeneracy}. The insights from this challenge provide key pointers to the important features of the next-generation of front-line imaging techniques of this type. Moreover, the data and evaluation routines remain available to form the basis of an expanding ranking of models and fitting routines and a standard yardstick for future model development.

The paper is organized as follows. We first describe in section \ref{sec:matandmet} the  experimental protocol,  data post-processing and preparation of the training and testing data. We then present the methods for ranking the models and tabulate succinctly the various models involved in the competition. We report the challenge results in section \ref{sec:results} and discuss these results in section \ref{sec:discussion}; a more detailed description of the models follows in the Appendix, section \ref{appendix}.

\section{\label{sec:matandmet}Material and Methods}

\subsection{The complete experiment protocol}
One healthy volunteer was scanned over two non-stop 4h sessions. The imaged volume comprised twenty 4mm-thick whole-brain sagittal slices covering the corpus callosum left-right. The image size was 110~x~110 and the in-plane resolution $2$~x~$2$~mm$^2$. Forty-five unique and evenly distributed diffusion directions (taken from http://www.camino.org.uk) were acquired for each shell, with both positive and negative polarities; these directions were the same in each shell. We also included $10$ interleaved b=$0$ measurements, leading to a total of $100$ measurements per shell. 
Each shell had a unique combination of $\Delta=\{22, 40, 60, 80, 100, 120\}$ ms, $\delta =\{3, 8\}$ ms, and $|{\bf G}|=\{60, 100, 200, 300\}$ mT/m (see Table \ref{fig:protocol}). The measurements were randomized within each shell, whereas the gradient strengths were interleaved. We visually inspected the images and have not observed any obvious shifts from gradient heating.
The minimum possible echo time (TE) for each gradient duration and diffusion time combination was chosen to enhance SNR for shorter diffusion times, and potentially enables estimation of compartment-specific relaxation constants. The SNR of b = 0 images was $35$ at TE = $49$ ms and $6$ at TE = $152$ ms. To find the SNR we used the background signal, as well as the signal noise floor in b = 0 images i.e. the residual signal along the fibres at the highest b-value. In both cases these estimates matched reasonably well. More details about the acquisition protocol can be found in \citet{ferizi_nimg}.

\subsection{Post-processing}
All post-processing was performed using FSL \citep{flirt}. The DW images were corrected for eddy current distortions separately for each combination of  $\delta$ and $\Delta$ using FSL's $Eddy$ module (www.fmrib.ox.ac.uk/fsl/eddy) with its default settings. 
The images were then co-registered using FSL's $Fnirt$ package. As the 48 shells were acquired across a wide range of TEs, over two days, we chose to proceed in two steps. First, within each quarter of the dataset (different day, different $\delta$) we registered all the $b$=0 images together. We then applied these transformations to their intermediary DW images, using a tri-linear resampling interpolation. The second stage involved co-registering the four different quarters. To help the co-registration, especially between the two days' images which required some through-plane adjustment as well, we omitted areas of considerable eddy-current distortions by reducing the number of slices from 20 to 5 (that is, leaving two images either side of the mid-sagittal plane) and reducing the in-plane image size to 75x80. 

\subsection{Training and testing data}
\begin{figure}[htb]
\begin{minipage}[b]{1.0\linewidth}
  \centering
  \centerline{\includegraphics[width=8cm]{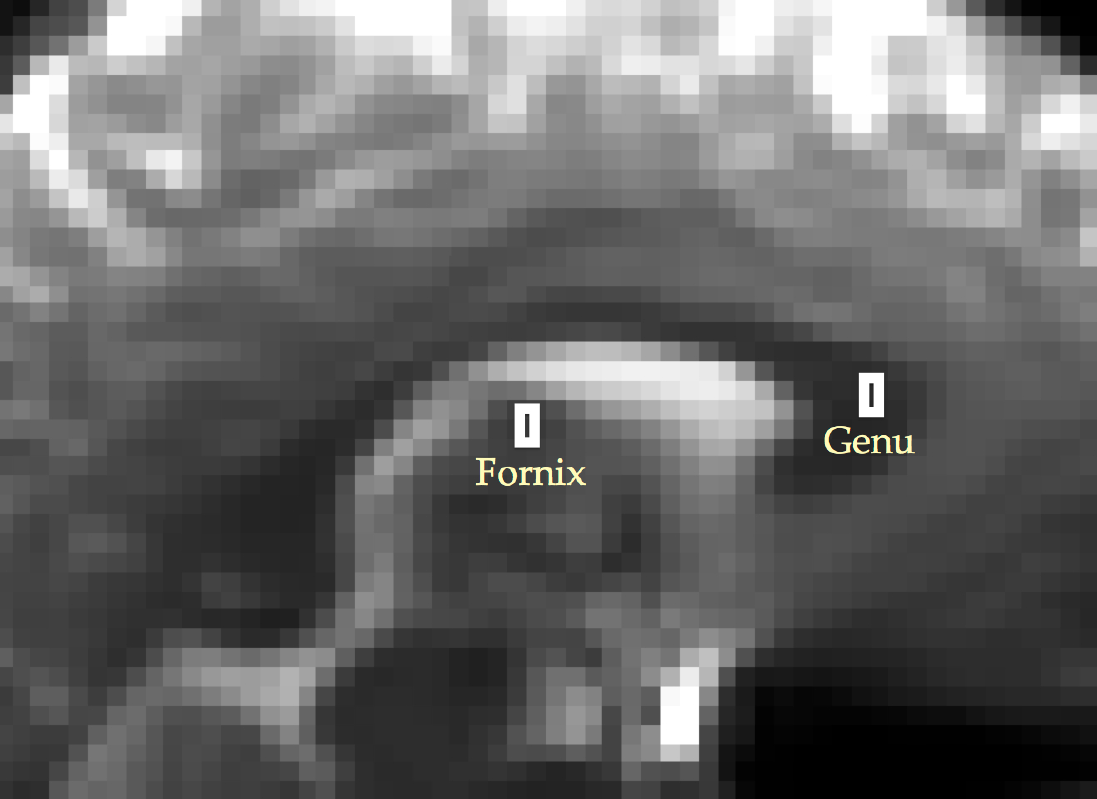}}
\end{minipage}
\caption[]{}
\caption[]{We only consider two ROIs, each containing six voxels from the genu in the corpus callosum, where the fibres are approximately straight and parallel, and from the fornix, where the configuration of fibres is more complex.}
\label{fig:roichoice}
\end{figure}

The data for this work originated from two ROIs, each containing 6 voxels (see Fig.~\ref{fig:roichoice}). The first ROI was selected in the middle of the genu in the corpus callosum, where the fibres are mostly  straight and coherent. The second ROI's fibre configuration is more complex: it lies in the body of fornix, where two bundles of fibers bend and bifurcate.

The dataset was split into two parts: the training dataset and the testing dataset. The training dataset was fully available for the challenge participants. The testing dataset was retained by the organisers. 
The DW signal of the training dataset (36 shells, with acquisition parameters shown in black in Table \ref{fig:protocol}) was provided together with the gradient scheme  on the challenge website (\url{http://cmic.cs.ucl.ac.uk/wmmchallenge/}). This data was used by the participants to estimate their DW-MRI model parameters.
The signal attenuation in the testing dataset (12 shells, 
with acquisition parameters shown in red in Table \ref{fig:protocol}) was kept unseen. Participants were then asked to predict the signal for the corresponding gradient scheme. 
The challenge participants were free to use as much or as little of the training data provided to predict the signal of the test dataset for the six voxels in each ROI.

\begin{figure}[htb]
\begin{minipage}[b]{1.0\linewidth}
  \centering
  \centerline{\includegraphics[width=16cm]{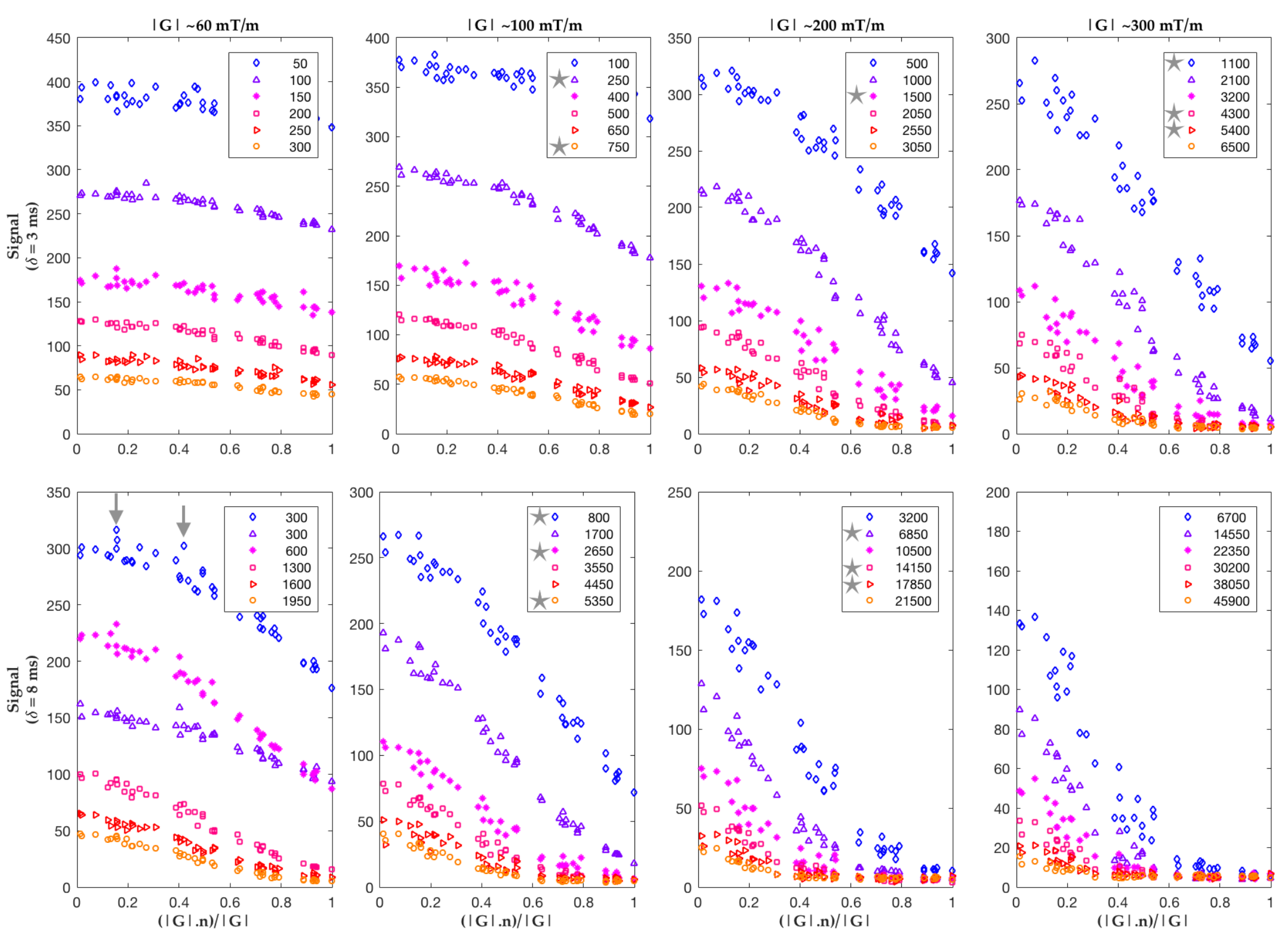}}
\end{minipage}
\caption[]{}
\caption[]{Diffusion weighted signal from the genu ROI, averaged over the six voxels. Across each column and row, the signal pertains to one of the gradient strengths or pulse times $\delta$ used; while in each subplot, the six shells shown in different colours are $\Delta$-specific, increasing in value  (22, 40, 60, 80, 100, 120 ms) from top to bottom. Inside the legend, the b-value is in s/mm$^2$ units; here, the HARDI shells kept for testing are those marked with a star; the remaining shells comprise the training data. On the x-axis is the cosine of the angle between the applied diffusion gradient vector \textbf{G} and the fibre direction \textbf{n}. Some models in this study omit data outliers; two such data points are shown in the bottom-left subplot with vertical arrows --- obviously each model has its own criteria for determining the outliers. }
\label{fig:signalPlot}
\end{figure}

\begin{figure}[htb]
\begin{minipage}[b]{1.0\linewidth}
  \centering
  \centerline{\includegraphics[width=16cm]{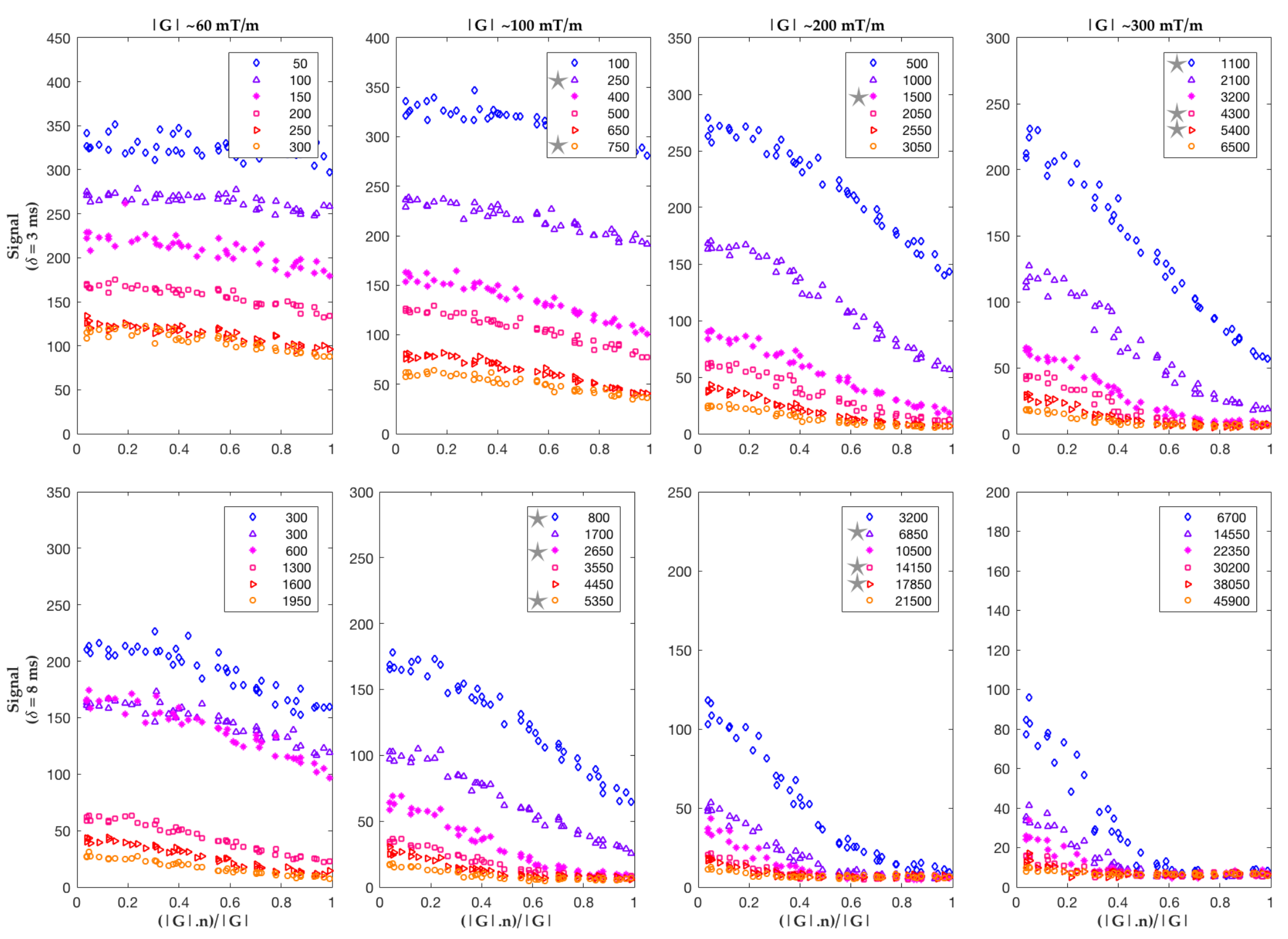}}
\end{minipage}
\caption[]{}
\caption[]{Similar to Figure \ref{fig:signalPlot}, here is the diffusion weighted signal from the fornix ROI, averaged over the six voxels.}
\label{fig:signalPlot2}
\end{figure}

Figure \ref{fig:signalPlot} shows the DW signal attenuation for each shell in the genu dataset, with stars in the legend indicating which shells were left out for testing.
In this plot, a small number of data appear as `outliers' (two such data are shown with arrows in the bottom-left subplot of Figure \ref{fig:signalPlot}). Specifically, we counted about 10 of them among more than 4812 measurements, most of them being in the b=300 s/mm$^2$ shell. Since these outliers appear to be specific to the b=300 s/mm$^2$ shell, and not in other shells with similar b-value, we attribute them to a momentary twitching of the subject rather than more systematic affects, such as perfusion.
Similarly, figure \ref{fig:signalPlot2} shows the signal for the fornix region, with the signal over the six voxels averaged out.

\subsection{Models ranking}
\label{ssec:modelRank}
Models were evaluated and ranked based on their ability to accurately predict the unseen DW signal. Specifically, the metric used was the sum of square differences between the hidden signal and the predicted signal, corrected for Rician noise \citep{jones2004squashing}:
\begin{equation}
SSE=\sum_{i=1}^N \frac{(\tilde{S}_i-\sqrt{S_i^2+\sigma^2})^2}{\sigma^2}
\label{eq:SSE}
\end{equation}
where $N$ is the number of measurements, $\tilde{S}_i$ is the $i$-th measured signal, $S_i$  its prediction from the model, and $\sigma$ is the noise standard deviation.

\subsection{Competing models}
\label{CompetingModels}
We give in this section a short summary of competing models in the challenge. Additionally, Table \ref{table:modelsummary} provides a summary of the key characteristics of the competing
models. A more detailed description of each model is included in the Appendix, section \ref{appendix}.
\begin{itemize}
\item \emph{Ramirez-Manzanares:} A dictionary-based technique that accounts for multiple fibre bundles, and models the distribution of tissues properties (axon radius, parallel diffusivity) and the orientation dispersion of fibres.
\item \emph{Nilsson:} A multi-compartment model that models isotropic, hindered and restricted diffusion, and accounts for varying  (T$_1$, T$_2$) relaxation times for each compartment \citep{nilsson}.
\item\emph{Scherrer} A multi-compartment model in which each compartment is modelled by a statistical distribution of 3-D tensors \citep{scherrer_mrm2015}
\item\emph{Ferizi$_1$ and Ferizi$_2$:} Two three-compartment models that account for varying T$_2$ relaxation times for each compartment. As regards the intracellular compartment, Ferizi$_1$  models the orientation dispersion by using dispersed sticks as one compartment; Ferizi$_2$ uses a single radius cylinder instead \citep{ferizi_nimg}.
\item\emph{Poot:} A 3-compartment model comprising an isotropic diffusion compartment, a tensor compartment, and a model-free compartment in which an ADC is estimated for each direction independently. T$_2$ relaxation times are also estimated for each compartment \citep{poot2014detecting}.
\item\emph{Rokem:} A combination of the sparse fascicle model \cite{Rokem2015} with restriction spectrum imaging \cite{White2013} that describes the signal arising from a multi-compartment model in a densely sampled spherical grid, using L1 regularization to enforce sparsity.
\item\emph{Eufracio:} An extension of the Diffusion Basis Function (DBF) model that accounts for multiple b-value shells.
\item\emph{Loyas-Olivas$_1$ and Loyas-Olivas$_2$:} Two models based on the Linear Acceleration of Sparse and Adaptive Diffusion Dictionary (LASADD) technique. Loyas-Olivas$_1$ uses the DBF signal model, while Loyas-Olivas$_2$ uses a three-compartment tissue model. The optimization uses linearized signal models to speed up computation and sparseness constraints to regularise. 
\item\emph{Alipoor:} A model of fourth-order tensors, corrected for T$_2$-relaxation across different shells. A robust LS fitting was applied to mitigate influence of outliers.
\item\emph{Sakaie} A two-compartment model of restricted and hindered diffusion with angular variation. A simple exclusion scheme based on the b=0 signal intensity was applied to remove outliers.
\item\emph{Fick:} A spatio-temporal signal model to simultaneously represent 3-D diffusion signal over varying diffusion time. Laplacian regularization was applied during the fitting \citep{fick2015unifying}.
\item\emph{Rivera:} A regularized linear regression model of diffusion encoding variables. This is intentionally built as a simplistic model to be used as a baseline for model comparison.   
\end{itemize}

\section{\label{sec:results}Results}
\begin{figure}[htb]
\begin{minipage}[b]{1.0\linewidth}
  \centering
  \centerline{\includegraphics[width=1.0\textwidth]{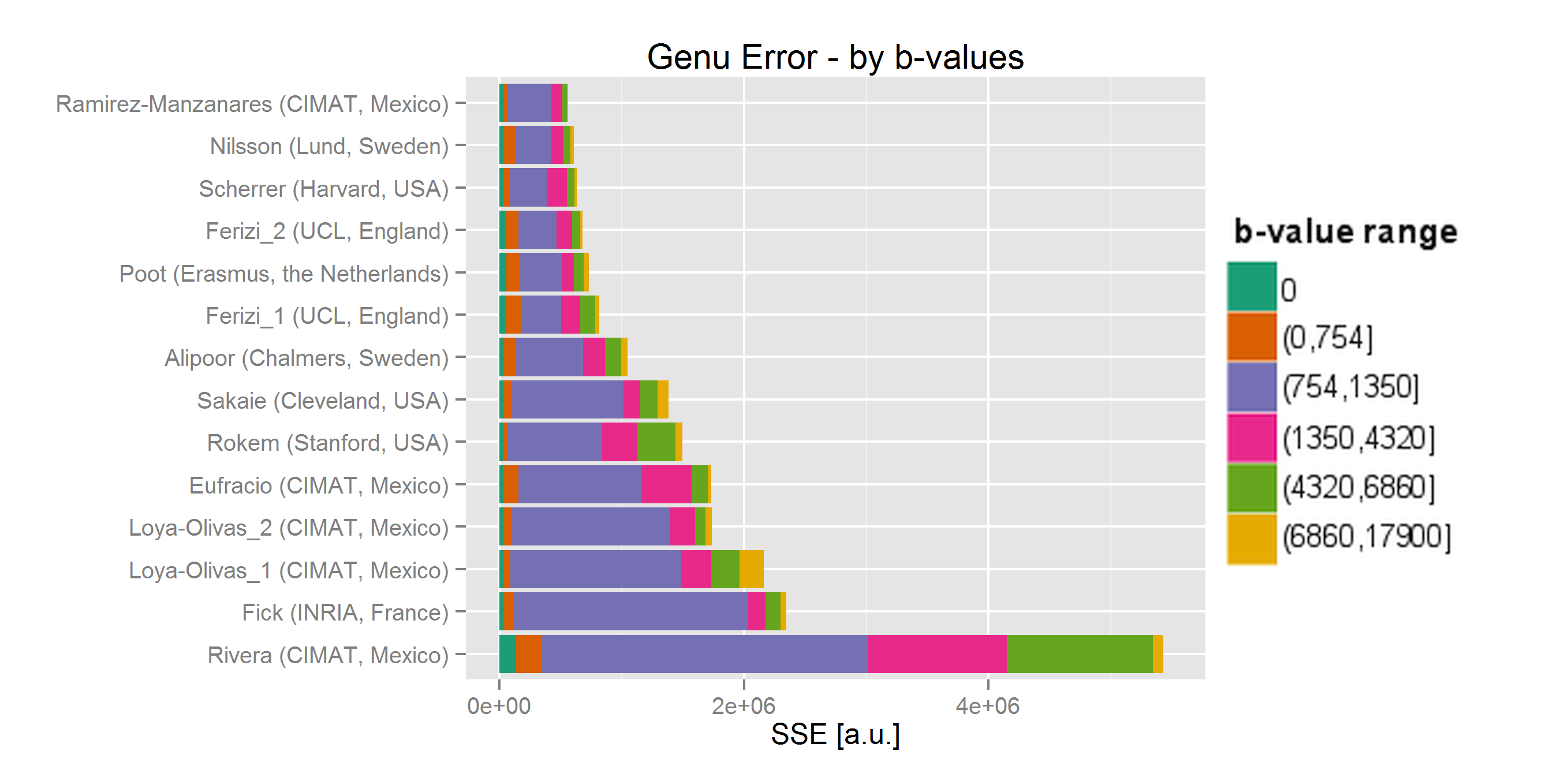}}
  \centerline{\includegraphics[width=1.0\textwidth]{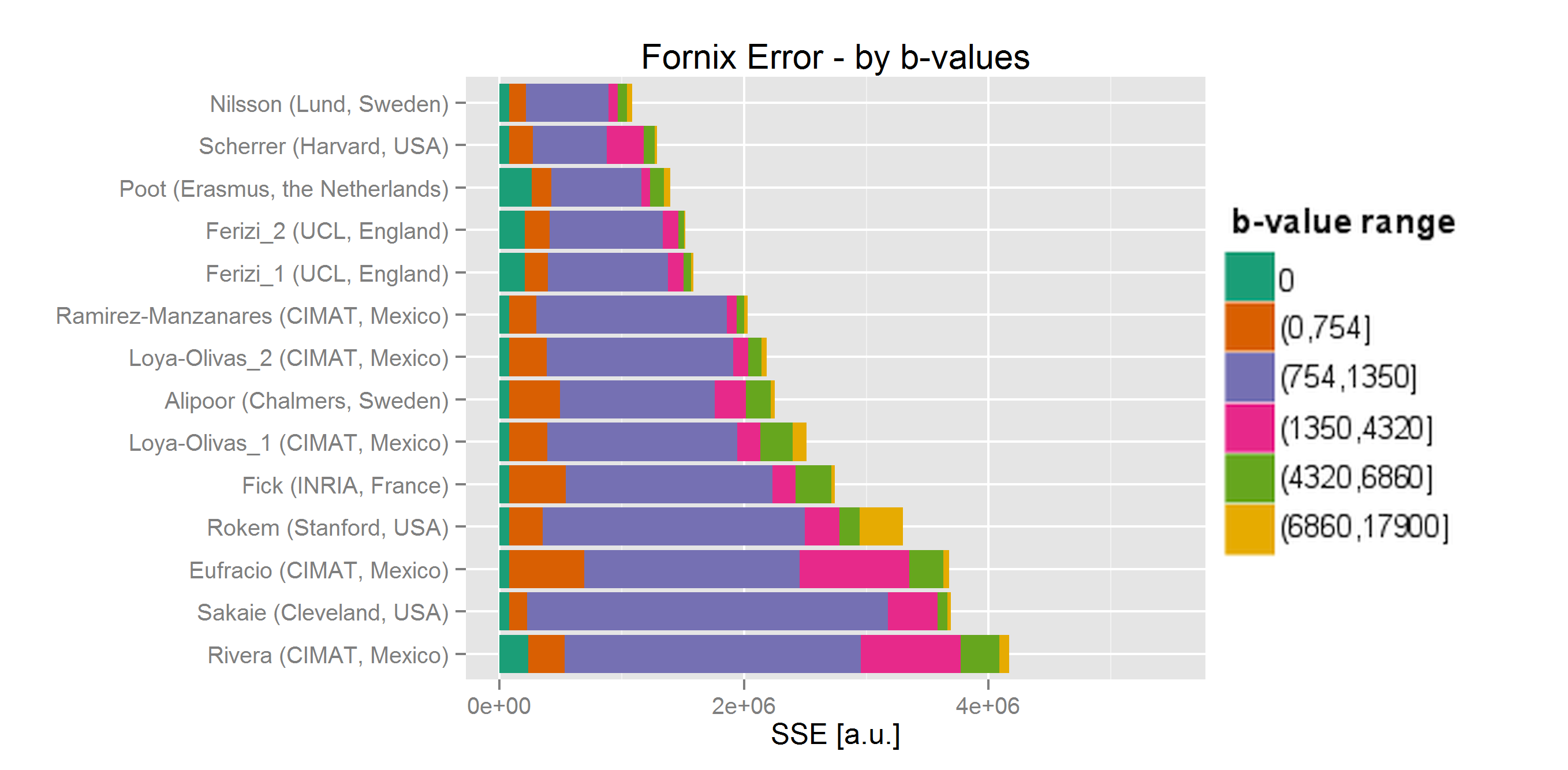}}
\end{minipage}
\caption[]{}
\caption[]{Overall ranking of models by sum-of-squared-error metric over all voxels in genu (top) and fornix (bottom) ROIs. The colors represent different ranges of b-value shells.}
\label{fig:rankingPlot}
\end{figure}

Figure \ref{fig:rankingPlot} shows the averaged prediction error in each ROI (top subplot is for the genu, bottom subplot is for the fornix) and the corresponding overall ranking of the participating models in the challenge.
The first six models in the genu ranking performed similarly, each higher ranked model marginally improving on the prediction error. The prediction error clearly increased at a higher rate for the subsequent models. 
In the fornix dataset, the prediction error was higher than in the genu.  For both datasets the first six models were the same, albeit permuted.
Most of the models performed similarly in terms of ranking in both genu and fornix cases, i.e. Nilsson (2nd in genu/1st in fornix), Scherrer (3rd/2nd) and Ferizi\_2 (4th/4th). Others performed significantly better in one of the cases, with Ramirez-Manzanares (1st/6th) being the most notable.

Figure \ref{fig:rankingPlot} also details the prediction error for different ranges of b-values in the unseen dataset. Models inevitably vary in their prediction capabilities; some models perform better within a given b-value range but are penalised more in another. Across the models, as the figure shows, the ranking between models was dominated by the signal prediction accuracy for b-values between 750 and 1400 s/mm$^2$; specifically, the shell which has the largest weight on this error is the b=1100 s/mm$^2$ one. 
The top-ranking models, nevertheless, were better at predicting the signal for higher b-value images as well. The prediction performance of lower b-value images (\textless750 s/mm$^2$) in the genu was less consistent across ranks. For example the models of Rokem and Sakaie outperformed most of the higher ranking models in this low b-value range. 
The fornix is a more complex region than the genu, hence the performance across the shells is less consistent. In the fornix the prediction errors were generally larger than in the genu across all $b$-values for all models, except Rivera's, which showed the opposite. The prediction errors of the $b=0$ images were also larger than in the genu, especially for the highly ranked models of Poot and Ferizi. The prediction errors in other $b$-value shells followed more closely the overall ranking of the models.

\begin{figure}[htb]
\begin{minipage}[b]{1.0\linewidth}
  \centering
  \centerline{\includegraphics[width=0.95\textwidth]{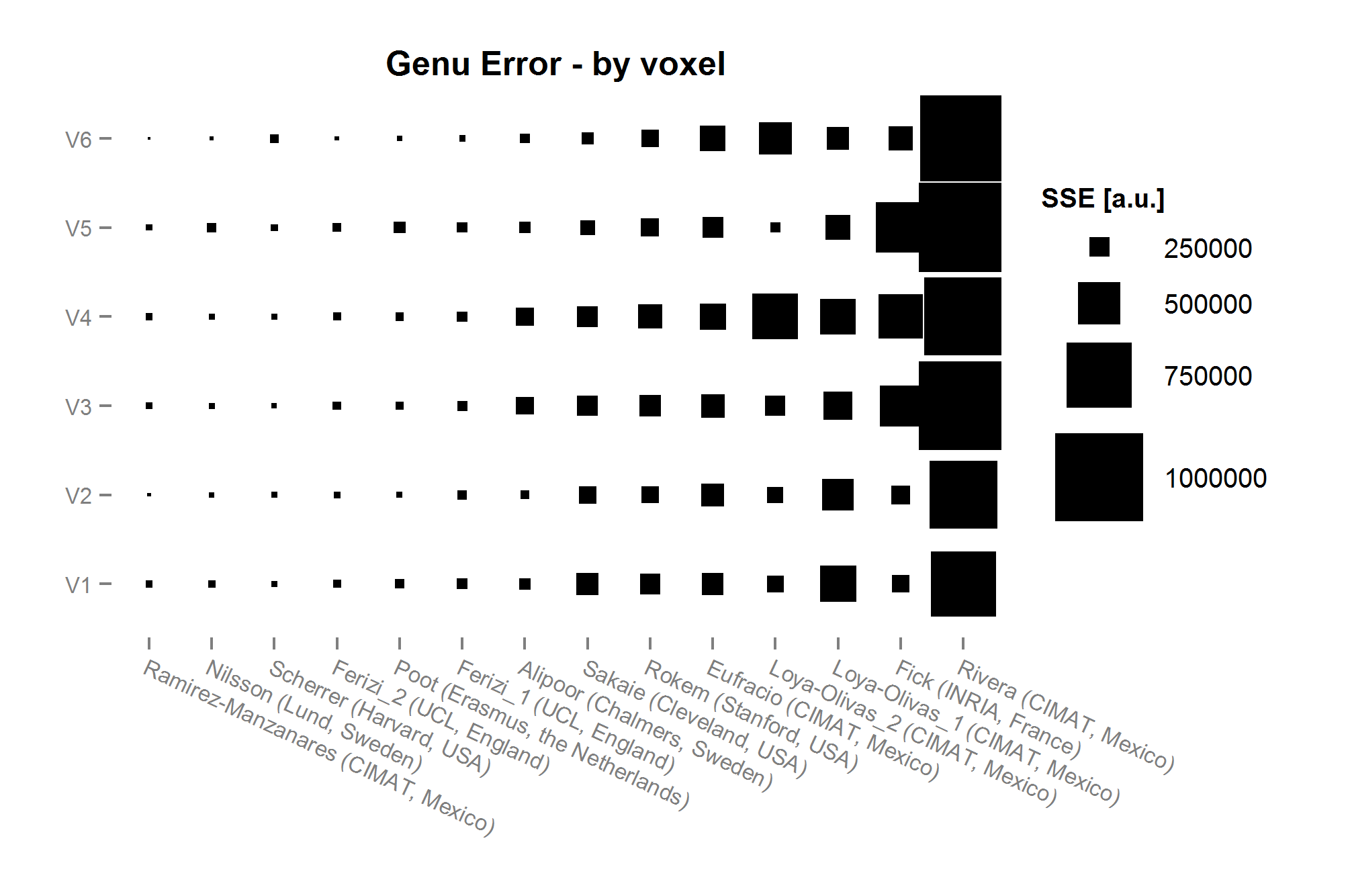}}
  \centerline{\includegraphics[width=0.95\textwidth]{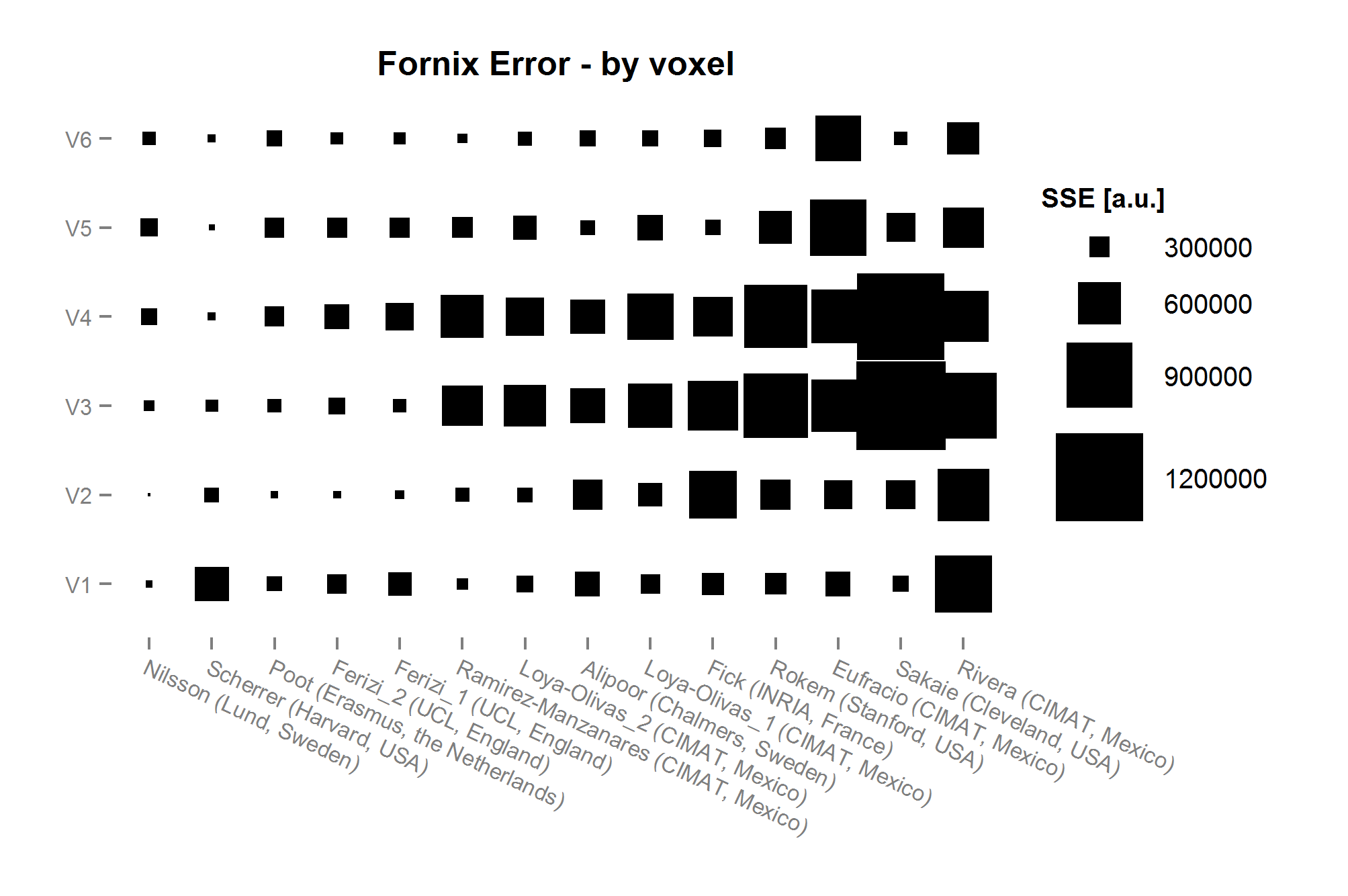}}
\end{minipage}
\caption[]{}
\caption[]{Sum-of-squared-error per voxel for each model in genu and fornix. The size of rectangles represent the SSE value per voxel.}
\label{fig:voxelErrors}
\end{figure}

Figure \ref{fig:voxelErrors} shows the prediction error for each voxel independently. In the genu plot, the best performing models had high consistency of low prediction errors across all individual voxels. These were followed by the models with consistent larger prediction error in all voxels.
Most of the lowest ranking models not only had largest prediction errors, they also showed large variations in prediction performance. For example, while the model of Loya-Olivas\_2 was competitive in voxel 5, it ranked low due to large prediction errors in voxel 4 and voxel 6.
The results in the fornix show a lower consistency of prediction errors between the voxels than in the genu. Specifically, two voxels (3 and 4) showed substantially larger prediction errors and were likely responsible for much of the overall ranking. 

\begin{figure}[htb]
\begin{minipage}[b]{1.0\linewidth}
  \centering
  \centerline{\includegraphics[width=13cm]{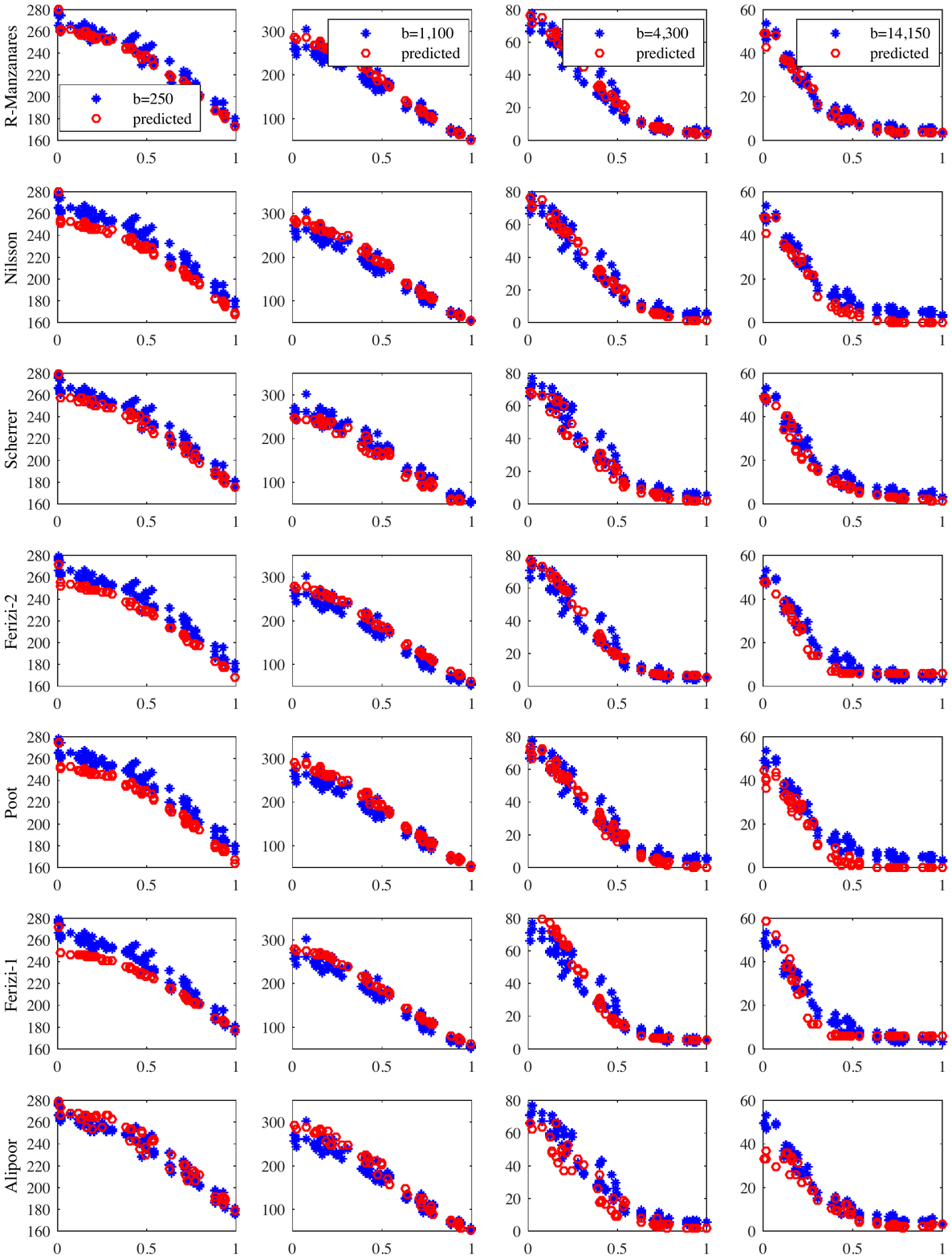}}
\end{minipage}
\caption[]{}
\caption[]{An illustration of the observed and predicted genu signal of 7 of the 14 best models, shown in red circles, to 4 (of the total 12) representative shells, shown in blue stars. The best models are listed first. The axes are as in Fig.\ref{fig:signalPlot}.}
\label{fig:illustration1}
\end{figure}

\begin{figure}[htb]
\begin{minipage}[b]{1.0\linewidth}
  \centering
  \centerline{\includegraphics[width=13cm]{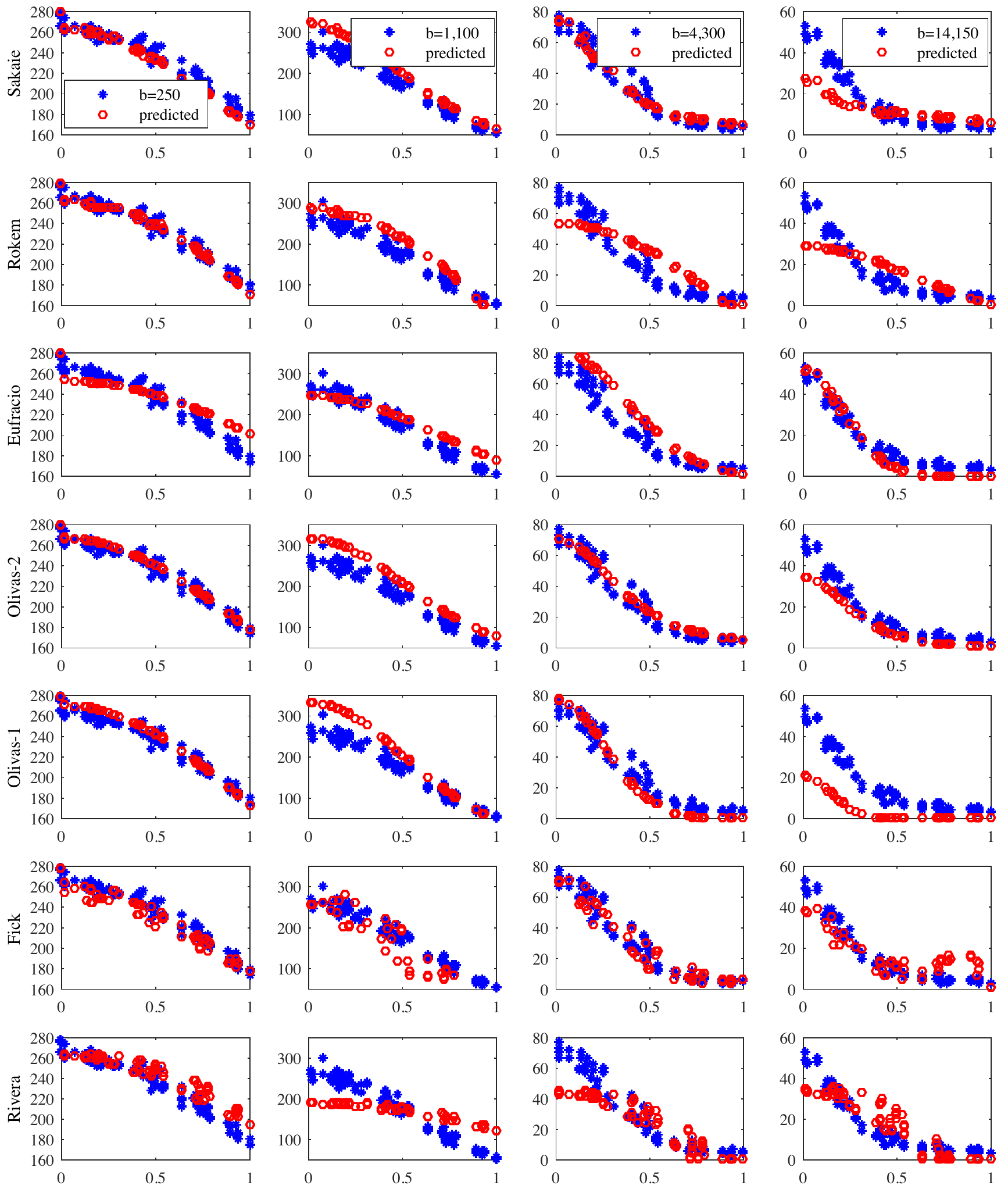}}
\end{minipage}
\caption[]{}
\caption[]{Similar to Fig.\ref{fig:illustration1}, here we show the remaining 7 model synthesized signals.}
\label{fig:illustration2}
\end{figure}

 Finally, we report in figures \ref{fig:illustration1} and \ref{fig:illustration2} an illustration of the quality of fit of each model to 4 representative shells, including the b=1100 s/mm$^2$ shell mentioned above.

\section{\label{sec:discussion}Discussion}
The challenge set out to compare the ability of various kinds of models to predict the diffusion MR signal from white matter over a very wide range of measurement parameters  -- exploring the boundaries of possible future quantitative diffusion MR techniques.
The fourteen challenge entries were a good representation of the many available models that are proposed in the literature. 
They also use a variety of fitting routines, and so provide additional insight into the effects of algorithmic choices during parameter estimation.
Although the set of methods tested is not sufficient to make a full comparison of each independent feature (diffusion model, noise model, fitting routine, etc.), and the number of combinations prohibits an exhaustive comparison, the results of the challenge do reveal some important trends.

\subsection{Main conclusions}
The first insight is on the type of model used. Signal models do not necessarily outrank tissue models; indeed, models of the signal (Alipoor, Sakaie, Fick, Riviera) ranked on average lower than models of the tissues with our dataset, despite their theoretical ability to offer more flexibility in describing the raw signal.
This is quite surprising, as the current perception within the field is that, generally, we can capture the signal variation much better through a functional description of the signal (signal models) rather than via a biophysical model of the tissue (tissue models). The former generally consist of bases of arbitrary complexity, whereas the latter are generally very parsimonious models that rely on extremely crude descriptions of tissue (e.g. white matter as parallel impermeable cylinders).
The results suggest that the flexibility of signal models can rapidly lead to over-fitting. 
However, the tissue models can explain the signal relatively well even with just a few parameters (compare the quality-of-fit plots of the Rivera model in figure \ref{fig:illustration2} to the signal prediction of the top models in figure \ref{fig:illustration1}; the higher the b-value, the worse the prediction of the linear signal model).
Certain underlying assumptions may cause the signal models to perform less well than expected. 
For example, they are often designed to work with data with a single diffusion time and do not generalise naturally to incorporate the additional dimension (although see \citet{fick2015unifying} for some steps towards generalisation). Many of the tissue models on the other hand naturally account for finite $\delta$, varying diffusion times and gradient strength (e.g. the Ramirez-Manzanares, Nilsson and Ferizi models in our collection).

The second insight concerns the choice of noise modelling. Assuming a non-Gaussian noise model, as used by three models, provides little benefit over a Gaussian assumption. This is likely because much of the data has high SNR. Although signal levels at high b-values do often hit the noise floor, the magnitude of the noise floor is small compared to signals at moderate b-values.
In this challenge most participants used non-linear least-squares or maximum likelihood optimisation. Additional regularisation of the objective function (Eufracio\&Rivera/Lasso, Rokem/Elastic Net, Fick/Laplacian) appeared to have little benefit over non-regularised optimization.

The third observation is about removing signal outliers. Five of the eleven models preprocessed the training data by clearing out outliers, including the top two models. We tried this procedure with two good models which did not use such a procedure, Ferizi$_1$ and Ferizi$_2$, and observed that it did not affect the ranking, though it did marginally improve the prediction error. This is understandable considering the relative little weight these apparent outliers have on the total number of measurements (10 points from a 4812-strong dataset). Additionally, specific strategies for predicting signal, e.g. bootstrapping or cross-validation, as used by the top two models of Ramirez-Manzanares and Nilsson, may also help the model ranking.

\subsection{Limitations and future directions}

Although this challenge provides several new insights into the choice of model and fitting procedure for diffusion-based quantitative imaging tools, it has a number of limitations that future challenges might be designed to address.
One limitations of the study is that we use a very rich acquisition protocol that is not representative of common or clinical acquisition protocols. In particular, we cover a very wide range of b-values and the data acquisition (protocol) we use consists of many TEs unlike many other multi-shell diffusion datasets that use a fixed TE. As stated in the Introduction, our intention is to sample the measurement space as widely as possible to support the most informative models possible. Varying the TE makes it possible to probe compartment-specific T$_2$ (whose decay \citet{ferizi_nimg} finds to be monoexponential at the voxel level), an investigation which would be impossible with a single TE. 
However, the good performance of DIAMOND also shows that a model with fixed $\delta$ and $\Delta$ could be used with multi-TE datasets, and that, while the majority of the full data was ignored in each of the reconstructions, its prediction error compared favourably with other techniques.

We use the unique human Connectome scanner \citep{connectom} to acquire a dataset with gradients of up to 300 mT/m, which is not readily available in most current MR machines. However, previous preclinical work by \citet{dyrby} suggests that high diffusion gradients enrich the signal, which helps model fitting and comparison. Future challenges might be designed that focus on explaining the signal and estimating parameters from data more typical of clinical acquisitions.

Assessing the prediction performance on unseen data as in this challenge is different from assessing the fitting error: it implicitly penalises models which overfit the data. However, since most of the missing shells lie in-between other shells (in terms of b-values, TEs, etc.), the quality of signal extrapolation was not assessed. 
We get a glimpse of this from figure \ref{fig:rankingPlot}, where the SSE is unevenly distributed between the b-values. Here, the shell which bore the largest error is the b=1100 s/mm$^2$ one; see also figures \ref{fig:illustration1} and \ref{fig:illustration2}. Of all ``unseen'' shells, this shell combines the lowest $\Delta$ and highest $|G|$, placing it on the edge of the range of the measurement space sampled. Such a b-value shell combines high signal magnitude with high sensitivity (i.e. the gradient of signal against b-value is highest in this range), which makes it hard to predict.
Future work can take this further, by selecting unseen shells outside of the min-max range of experimental parameters. This is likely to penalise more complex models that overfit the data even more strongly.

We did not take into account the computational demand of each model, and this might limit the generalisation of the results. Models that use bootstrapping generally have a higher computational burden and may not be feasible for large datasets, e.g. whole brain coverage.  

The dataset used in this challenge is specific to one subject who underwent a long duration acquisition, which adds to the question of the generalisability. The subsequent preprocessing of the data is also a factor to bear in mind: the registration of two 4h datasets, across such a broad range of echo times, poses its own challenges for certain non-homogenous regions in the brain, such as the fornix (as compared with, for example, the relatively large genu). Thus the results may be somewhat subject specific and may be affected by residual alignment errors. 

Another limitation is that we only look at isolated voxels inside the corpus callosum and the fornix. Questions still remain about which models are viable even in the most coherent areas of the brain with the simplest geometry so we believe our focused challenge on well-defined areas is an informative first step necessary before extending the idea to the whole of white matter, which would make for an extremely complex challenge. We note however, recent work by \citet{ghosh2016} that illustrates such an approach with Human Connectome Project (HCP) data.

We focused here on comparing models based on their ability to predict unseen data. Although models that reflect true underlying tissue structure should explain the data well, we cannot infer in general that models that predict unseen data better are mechanistically closer to the tissue than those that do not.
As we discuss in the introduction, the main power of evaluating models in terms of prediction error is to reject models that cannot explain the data. Thus, while the identification of parsimonious models that explain the data certainly has great benefit, further validation is necessary through comparison of the parameters that they estimate with independent measurements, e.g. obtained through microscopy (our challenge makes no attempt to assess the integrity of parameter estimates themselves, but future challenges might use such performance criteria).
We note however that major difficulties arise in obtaining ground truth in realistic samples. In particular, histology does not provide a perfect ground truth for assessing quantitative non-invasive techniques. There are two good reasons for this: a) the histological preparation process disrupts the tissue from its in-vivo state and b) certain parameters simply cannot be measured histologically, e.g. diffusivity and permeability. Moreover, the results from this study do not immediately translate into the ability of estimated models to provide useful information about the WM microstructure integrity, such as the presence of axonal loss, demyelination or oedema in abnormal tissue. Specifically for tissue models, obtaining a direct histological analogue is often difficult. One good example of this are models that incorporate an axonal radius parameter, known to generally overestimate the true axonal radius, as discussed in detail e.g. by \citet{barazany}, \citet{activeax} and \citet{dyrby}.

\section{Conclusion}
Challenges such as this have great value in bringing the community together and provide unbiased comparison of wide ranging solutions to key data-processing problems. They raise new insights and ideas motivating more directed future studies. The data is publicly available for others to use, with more details of the dataset given on the Challenge website http://cmic.cs.ucl.ac.uk/wmmchallenge/. On this website, an up-to-date ranking of the models will be available too, where additional models can be added after the publication of the article. This will provide an important yardstick for future models and parameter estimation routines.



\section{Acknowledgements}
\label{sec:ack}
Research reported in this manuscript was supported by:
\begin{itemize}
\item EPSRC supported this work through grants  EP/G007748, EP/L022680/1,  EP/I027084/01, EP/M020533/1 and EP/N018702/1.
\item U. Ferizi is also supported by the National Institute of Arthritis and Musculoskeletal and Skin Diseases (NIAMS) of the National Institute of Health (NIH) under award numbers R21AR066897 and RO1AR067789. 
\item B. Scherrer was supported in part by NIH R01 NS079788, R01 EB019483, R01 EB018988 and BCH TRP Pilot and BCH CTREC K-to-R Merit Award.
\item M. Nilsson is supported by the Swedish Strategic Research (SSF) Grant AM13- 0090
\end{itemize}
The content is solely the responsibility of the authors, and does not necessarily represent the official views of the funding bodies (EPSRC or NIH).

Lastly, we would like to thank ISBI 2015 challenge organisers, in particular, Stephen R. Aylward (Kitware Inc., USA) and Badri Roysam (University of Houston, USA).

\section{Appendix: Competing models} 
\label{appendix}

\subsection{Tissue models}
\subsubsection{Ramirez-Manzanares (CIMAT, Mexico): Empirical Diffusion-and-Direction Distributions (ED$^3$)} 
This work builds on the statistical modelling of the apparent diffusion coefficient \citep{yablonskiy2003}, and tackles the modelling of axon fiber dispersion in single \citep{Axer2001,zhang_noddi} and multiple fibre bundle cases. The method empirically estimates (rather than imposes)  the distribution of tissue properties (axon radius, parallel diffusion, etc.), as well as the orientational distribution of the bundles. The general framework is as follows: 
\begin{itemize}
\item estimation of mean principal diffusion directions (PDD) per axon bundle;
\item selection of a dense set of orientationally--focused basis directions that capture the discrete non--parametric fiber dispersion; 
\item design of a dictionary of  intra/extra cellular synthetic DW--signals which are precomputed along the basis directions (see the DBF method in \citet{ramirez-manzanares2007}); 
\item computation of the size compartments per diffusion atom of the dictionary (model fitting). 
\end{itemize}

The PDDs are estimated from the DT (single bundle case) and DBF \citep{ramirez-manzanares2007} (complex structure cases).
The 120 orientations closest to the PDDs are selected from a set of 1000, evenly distributed orientations. The intra axonal signals $S^{i}$ are precomputed from the model in \citet{van_gelderen94} for restricted diffusion within a cylinder with radius $R= 1, 2, \dots, 10$ ${\mu}m$ and parallel diffusion $d_{\parallel} = 1, 1.1,\dots, 2.1$ ${\mu}m^2/ms$. The extra-axonal signals are generated as: $S^{e}$ from  \textit{zeppelins} with combinations of parallel  and radial diffusion, $d_{\parallel} = 1, 1.1,\dots, 2.5$ ${\mu}m^2/ms$ and $d_{\perp} = 2, 3,\dots, 8$ ${\mu}m^2/ms$, the isotropic diffusion compartment signals $S^{iso}_i = \exp^{-q \tau d_{iso} }$ for $d_{iso} = 2, 2.1, 2.2, \dots , 4$ ${\mu}m^2/ms$, and the \textit{dot} signal that takes into account static proton density. The values of the dictionary-atoms above were tuned by cross validation \citep{crossvalidation}. The size compartments $\beta \geq 0$ computed in the weighted non--negative LS formulation:
\begin{equation}
 \left | \left | W\left(S - S_0^{TE}\left(  \sum_{i=1}^{N_{i}} \beta_i^{i} S_i^{i} -  \sum_{j=1}^{N_{e}} \beta_j^{e} S_j^{e} -\sum_{k=1}^{N_{iso}} \beta^{iso}_k S_k^{iso} + \beta_{dot} \right)\right) \right | \right |_2^2
\label{ramirezEQ1}
\end{equation}

  indicate the atoms that explain the signal, the $W$ weights are proportional to SNR. Overfitting is reduced by a \textit{bootstrap} \citep{bootstrap}  procedure.

The cross-validation experiments indicate that the reconstructions given by the robust fitting of this rich multi-compartment diffusion dictionary allows to accurately predict non acquired MR signals for different machine protocols.  This is of most interest  in the development of methods able to detect the complex microstructure heterogeneity associated with the different compartments within the voxels. The atoms with coefficients $\beta>0$  depict the empirical distributions, and their orientations indicate non-parametrical bundle--dispersion configurations (as fanning or radially symmetrical). The recovered distributions reveal, for instance, the presence of axon radius only among 1 and 4 $\mu m$. One should take into account, however, that since the heterogeneous intra/extra-axonal T2 relaxation feature is not explicitly modeled, the method may compensate T2 variations by using, for instance, large isotropic $d_{iso}$ coefficients to accurately fit the signal. For this reason, a direct interpretation of the fitted parameters may be misleading. The use of more specific models is a part of ongoing work.

\subsubsection{Nilsson (Lund, Sweden) : Multi-compartment model outlier rejection and separate fitting of b0 data} 
This multiple compartment model was developed specifically for the ISBI WM challenge and built up by a relaxation-weighted and time-dependent diffusion tensors according to 
\begin{equation}
S = S_0 \sum_i w_i~e^{-\mathbf{B}:\mathbf{D}_i} e^{\rm -TE/T2_i} \Big(1 - e^{\rm -(TR - TE/2)/T1_i} \Big)
\label{eq:mn_model}
\end{equation}
where $\mathbf{B} = b \vec{n}^{\otimes 2}$ and $b = (\gamma \delta g)^2 t_d$. The diffusion time $t_d$ was corrected for rise times ($\xi$) according to $t_d = \Delta - \delta/3 + \xi^3/30\delta^2 - \xi^2/6\delta$. Each component was also described by a weight ($w_i$) and relaxation times (T1$_i$ and T2$_i$). The model featured three types of components, with either isotropic, hindered and restricted diffusion. Diffusion in the isotropic component was modeled by a single diffusion coefficient. The hindered and restricted components were modeled by cylinder-symmetric tensors described by axial and radial diffusivities together with the polar and azimuth angles. In the restricted component, the apparent diffusion coefficient of the radial component depended on $\delta$ and $\Delta$, as well as on the cylinder radius, according to \citet{vangelderen}.

Three modifications were performed to this very general model. First, to accommodate for potential bias in the $b_0$ images (which was the case for fornix data where deviations of up to 20$\sigma$ was observed), the prediction for $b_0$ data was obtained from the median of all signals acquired with identical TE instead of from eq. \ref{eq:mn_model}. Second, opposite direction acquisitions were rescaled by a free model parameter, in order to allow for potential gradient instabilities inducing differences between the directions and their opposite directions. Third, models were generated dynamically during fitting by randomly selecting up to four hindered components and up to three restricted components. One isotropic component was always included. 

The model was first fitted to half of the diffusion-weighted data (randomly selected), after which outliers were rejected ($>2.5\sigma$). Thereafter a second fit was performed. Both fit steps assumed Gaussian noise and utilized the 'lsqcurvefit' function in Matlab. The procedure was repeated 100 times for different randomly generated models. 

To prepare for submission of the results, only the models that best predicted the hidden half of the data was selected, after which the median of the selected predictions were used for the final prediction.

\subsubsection{Scherrer (Harvard, USA): Distribution of anisotropic microstructural environments in diffusion compartment imaging (DIAMOND)} 
DIAMOND models the set of tissue compartments in each voxel by a finite sum of unimodal continuous distributions of diffusion tensors. This corresponds to a hybrid tissue model that combines biophysical and statistical modeling. 
As described in \citep{scherrer_mrm2015}, the DW  signal $S_k$ for a gradient vector $\mathbf{g}_k$ and b-value $b_k$ is modeled by: 
$
S_k = S_0 \big[  \sum_{j=0}^{N} f_j  {\left( 1+ \frac{b_k \mathbf{g}_k^T  \mathbf{D}^0_j  \mathbf{g}_k }{\kappa_j}  \right)}^{-\kappa_j} \big]$,
where $S_0$ is the non-attenuated signal, $N$ is the number of compartments, $f_j$ the relative fraction of occupancy of the j$^\mathrm{th}$ compartment and $\kappa_j$ are $ \mathbf{D}^0_j$  are respectively the concentration and the expectation of the j$^\mathrm{th}$ continuous tensor distribution.
DIAMOND enables  assessment of compartment-specific diffusion characteristics such as the compartment FA (cFA), the compartment RD (cRD) and the compartment MD (cMD). It also provides a novel measure of microstructural heterogeneity for each compartment. 

The estimation of a continuous distribution of diffusion tensors requires DW data acquired with same timing parameters $\delta$ and $\Delta$ \citep{scherrer_mrm2015}. To compare DIAMOND to other models on this dataset, we fitted separately one DIAMOND model for each \{$\delta$, $\Delta$\} group  (\textit{i.e.}, for each TE group), leading to $12$ DIAMOND models. One shell was missing in each TE group; we predicted its signal using the corresponding DIAMOND model. The model estimation was achieved as follows. We first computed the mean  and standard deviation  of $S_0$ ($\mu_{S_0}$ and $\sigma_{S_0}$) within each TE group and discarded DW-signals whose intensity were larger than $\mu_{S_0} + 3\sigma_{S_0}$ (simple artefact correction). We then estimated DIAMOND parameters as described in \citet{scherrer_mrm2015}, considering Gaussian noise and cylindrical anisotropic compartments. 
For the genu we considered a model with one freely diffusing and one anisotropic compartment; for the fornix we considered a model with  one freely diffusing compartment and two anisotropic compartments.

\subsubsection{Ferizi\_1 and Ferizi\_2 (UCL, England)}
This submission uses two three-compartment models, as described in previous studies \citep{ferizi_mrm,ferizi_miccai}. These models consist of: 1) either a Bingham distribution of sticks or a Cylinder for the intracellular compartment; 2) a diffusion tensor for the extracellular compartment; 3) an isotropic CSF compartment. 
The T$_2$ relaxation element is fitted beforehand, to the (variable echo time) b=0 measurements. The signal model is:  
\begin{equation}
S = \tilde{S_{0}} \left(f_{i}  \exp{(-\frac{TE}{T_2^i})}  S_{i} + f_{e}  \exp{(-\frac{TE}{T_2^e})}  S_{e} + f_{c}  \exp{(-\frac{TE}{T_2^c})}  S_{c}\right)
\label{eq:modelTE}
\end{equation}
where $f_{i}$, $f_{e}$ and $f_{c}$ are the weights of the intracellular, extracellular, and third normalised compartment signals $S_{i}$, $S_{e}$ and $S_{c}$, respectively; the values of compartmental $T_2$ are indexed similarly; $\tilde{S_{0}}$ is the proton density signal (which is TE-independent, and obtained from fitting to the b = 0 signal).
These models, as shown in the figure below,  emerged from previous studies (see references below). Here, however, a single white matter T2 and separate compartmental diffusivities are additionally fitted.

There is a two-stage model fitting procedure. The first step estimates the T2 decay rate of tissue, separately in each voxel, by fitting a bi-exponential model to the b=0 intensity as a function of TE, in which one component is from tissue and the other from CSF. A preliminary analysis of voxels fully inside WM regions shows no significant departure from mono-exponential decay, equal T2 are then assumed within the intra and extracellular compartments. When fitting the bi-exponential model, the value of T2 in CSF is fixed to 1,000ms (a more precise value of CSF is unlikely to be estimated with this protocol). Thus, for each voxel, the volume fraction of CSF, the $\tilde{S_{0}}$ and the T2 of the tissue are estimated. These three estimates are then fixed for all the subsequent model fits. Then, each model is fitted using the Levenberg-Marquardt algorithm with an offset-Gaussian noise model.

\subsubsection{Poot (Erasmus, the Netherlands)} 
This submission uses a three compartment model, with for each compartment a different complexity of the diffusion model and an individual $T_2$ value. This model was developed specifically for the ISBI WM challenge and is the result of iteratively visualizing different projections of the residuals and trying to infer the maximum complexity that the rich data supports. 

The first compartment models isotropic diffusion and, through the initialization procedure, it captures the fast diffusion components. The second compartment is modelled by a second order (diffusion) tensor and models intermediate diffusion strengths. The third compartment is model-free as the ADC is estimated for each direction independently. Each compartment additionally has an individual $T_2$ value and signal intensity at $b=0$, $\TE=0$ (which could easily be translated into volume fractions). Hence, the complete model of a voxel in image $j$ is given by 
\begin{align}
  S_j(\bm \theta) = \sum_{i=1}^3 A_i e^{-\TE_j R_{2,i}} e^{-b_j \ADC_{j,i} } = \sum_{i=1}^3 e^{\bm M_{i,j} \bm \theta} \label{EQ:Poot_S} 
\end{align}
where $S_j$ is the predicted signal intensity of image $j$, $A_i$ is the non-diffusion weighted signal intensity of compartment $i$ at zero $\TE$, $\TE$ is the echo time, $R_2$ is the reciprocal of the $T_2$ relaxation time of compartment $i$, $b=(\Delta - \delta/3) \delta^2 |G|^2 \gamma^2$, with $\gamma = 42.5781 \MHz/T$, $\ADC_{j,1} = c$, $\ADC_{j,2}= \bm g_j^T \bm D \bm g_j$, $\ADC_{j,3} =\bm d  \bm h^T_j$, where $\bm d$ is a vector with the ADC value of each orientation group and $\bm h_j$ is a vector that selects the orientation group to which image $j$ belongs (90 groups in total). Note that $\bm h_j$ has at most one nonzero element and that element has a value of one.  
As displayed in the right most part of \refeq{EQ:Poot_S}, the model can be written as a multiplication of matrices $\bm M_{i}$, containing all rows $\bm M_{i,j}$, with $\bm \theta = [\ln A_1,\  R_{2,1},\   c,\ \ \ln A_2,\  R_{2,2},\  D_{11},\  D_{12},\  D_{13},\  D_{22},\  D_{23},\  D_{33},\  \ \ln A_3,\  R_{2,3},\  \bm d]^T$, which combines all $103$ parameters into a single parameter vector. All parameters are simultaneously estimated from the provided $3\,311$ measurements per voxel by a maximum likelihood estimator that assumes a Rician distribution of the measurements and simultaneously optimizes the noise level \citep{poot2014detecting}. The exact initialization and details of the optimization procedure are provided in the online supporting material. Finally, the signal intensities of the `unseen' data are predicted by substituting the estimate into \refeq{EQ:Poot_S}.

\subsubsection{Rokem (Standford, USA): A restriction spectrum sparse fascicle model (RS-SFM)}

The Sparse Fascicle Model, SFM \citep{Rokem2015}, is a member of the large family of models that account for the diffusion MRI signal in the white matter as a combination of signals due to compartments corresponding to different axonal fiber populations (fascicles), and other parts of the tissue. Model fitting proceeds in two steps. First, an isotropic component is fit. We model the effects of both the measurement echo time (TE), as well as the measurement b-value on the signal. These are fit as a $log(TE)$-dependent decay with a low order polynomial function, and a b-value-dependent multi-exponential decay (including also an offset to account for the Rician noise floor). The residuals from the isotropic component are then deconvolved with the perturbations in the signal due to a set of fascicle kernels each modeled as a radially symmetric ($\lambda_2=\lambda_3$) diffusion tensor. The putative kernels are distributed in a dense sampling grid on the sphere. Furthermore, Restriction Spectrum Imaging (RSI \citep{White2013}) is used to extend the model, by adding a range of fascicle kernels in each sampling point, with different axial and radial diffusivities, capturing diffusion at different scales. To restrict the number of anisotropic components (fascicles) in each voxel, and to prevent overfitting, the RS-SFM model employs the Elastic Net algorithm (EN \citep{Zou2005}), which applies a tunable combination of L1 and L2 regularization on the weights of the fascicle kernels. We used elements of the SFM implemented in the dipy software library \citep{Garyfallidis2014} and the EN implemented in scikit-learn \citep{pedregosa2011}. In addition, to account for differences in SNR, we implemented a weighted least-squares strategy whereby each signal's contribution to the fit was weighted by its TE, as well as the gradient strength used. EN has two tuning parameters determining: 1) the ratio of L1-to-L2 regularization, and 2) the weight of the regularization relative to the least-squares fit to the signal. To find the proper values of these parameters, we employed k-fold cross-validation \citep{Rokem2015}, leaving out one shell of measurement in each iteration for cross-validation. We determined that the tuning parameters with the lowest LSE \citep{panagiotaki} provide an almost-even balance of L1 and L2 penalty with weak overall regularization. Because of the combination of a dense sampling grid (362 points distributed on the sphere), and multiple restriction kernels (45 per sampling point), the maximal number of parameters for the model is approximately 16,300, more than the number of data points. However, because regularization is employed, the effective number of parameters is much smaller, resulting in an active set of approximately 20 regressors \citep{Zou2007}. We have made code to fully reproduce our results available at \url{https://arokem.github.io/ISBI2015}.

\subsubsection{Eufracio (CIMAT, Mexico): Diffusion Basis Functions for Multi--Shell Scheme} 

This model is based on the Diffusion Basis Functions (DBF) model \citep{ramirez-manzanares2007}, a discrete version of the Gaussian Mixture Model for the sphere: $\hat s_i = \sum_{j=1}^m \alpha_j  \phi_{ij} + \epsilon$, with $\hat s_i=s_i/s_0$, $\phi_{ij} =  \exp \big(-bq_i^TT_jq_i \big)$  and $T_j = (\chi_1 v_jv_j^T + \chi_2I)$. The DBF model is reformulated by substituting $\phi_{ij}$ and $T_j$: $\hat{s}_i = \sum_{j=1}^m \alpha_j \exp \left(- b_i \chi_2 g_i^T g_i \right) \exp \left( -b_i \chi_1 (v_j^T g_i)^2 \right) + \epsilon$. The first exponential can be defined as a scale factor that depends on the b-values, $\beta_{i} = \exp ( -b_i \chi_2 q_i^Tq_i )$. In this way, the $\beta_i$ factors are associated with different b-values, so the new model includes the information of multi-shell schemes. The coefficients $\alpha$ and the shell scale factor $\beta$  are computed by solving the optimisation problem:

\begin{equation}
\begin{aligned}
\label{eq:newcost}
& \underset{\alpha,\beta_c}{\text{min}} & & f(\alpha, \beta_c; \lambda_{\alpha},\lambda_{\beta} )= \| B \tilde \Phi \alpha - S \|^2_2 + \lambda_\alpha \|\alpha\|_1  + \lambda_{\beta} \| \beta_c^0 - \beta_c\|^2_2 \; \; \; \text{s.t.} \; {\bf 1}^T\alpha = 1, \alpha \geq 0
\end{aligned}
\end{equation}

\noindent where $B = \operatorname{diag}(\beta_c)$,  $ \beta_c = \frac{1}{\#C}\sum_{i \in C}  \exp \Big( -b_i \hat \chi_2(q_i^Tq_i)\Big)$ and $C$ is the set of indices grouped by different b-values (and \#C is the number of elements in it). The regularization term  weighted by $\lambda_\alpha$ demands sparseness and the term weighted by $\lambda_{\beta}$ prevents an over-fitting.  The problem in eq.\eqref{eq:newcost} is solved in three steps. First, the active atoms are predicted ($\alpha_i > 0$) with $\tilde \alpha = \operatorname{argmin}_{\alpha} f(\alpha, \beta_c; \lambda_{\alpha},\lambda_{\beta} )$. Second, the active atoms are corrected with $ \alpha = \operatorname{argmin}_{\{\alpha_i\} : \tilde \alpha_i >0} f(\alpha, \beta_c; 0,\lambda_{\beta} )$. Finally,  the factors $\beta_c$ are updated with $\beta_c = \operatorname{argmin}_{\beta_c} f(\alpha, \beta_c; \lambda_{\alpha},\lambda_{\beta} )$. To solve each step, the active sets algorithm for quadratic programming is used.

To train the model for the WMM'15 data, eq.\ref{eq:newcost} is solved for each voxel with the training data to find the optimal weights $\alpha_j$ and scale factors $\beta_c$ that best reproduce the training data. For this challenge, the $\beta_c$ factors are grouped by the 36 training shells and the method parameters are set by hand: $\lambda_\alpha = 0.5$, $\lambda_{\beta} = 0.02$, $\chi_1 = 9.5 \times 10^{-4}$and $\chi_2 = 5 \times 10^{-5}$. To predict the unseen signal at each voxel, the reformulated model is used with the optimal weights $\alpha_j$ and the 12 scale factors for the unseen $\beta_c$ are calculated by interpolation with the 36 optimal $\beta_c$ of the training data. 

\subsubsection{Loya-Olivas\_1 and Loya-Olivas\_2 (CIMAT, Mexico): Linear Acceleration of Sparse and Adaptive Diffusion Dictionary (LASADD)}

\noindent LASADD is a multi--tensor based technique to adapt dynamically the  Diffusion Functions (DFs) dictionary to a DW--MRI signal \citep{lasadd_ismrm,loya_masterthesis}. The method changes size and orientation of relevant Diffusion Tensors (DTs). The optimisation algorithm uses a special DT expression and assumptions to reduce the computational cost.

The one--compartment version (LASADD--1C) is based on DBF multi--tensor model \citep{ramirez-manzanares2007}:
$s_i^{*} = \sum_{j=1}^n \alpha_j \phi_{i,j}$ where $s_i^{*}=\frac{s_i}{{s_0}_i}$, $\phi_{i,j} = \exp \left( -b_i \mathbf{g}_i^T \mathbf{T}_j \mathbf{g}_i \right)$, $\alpha_j>0$, and $\sum_{j=1}^n \alpha_j=1$. LASSAD expresses the DT as
\begin{equation}
\label{lasadd_tensor}
 \mathbf{T}_j={\chi_1}_j \mathbf{v}_j \mathbf{v}_j^T + {\chi_2}_j \mathbf{I},
\end{equation}
where ${\chi_{\{1,2\}}}_j$ are scalars associated to the eigenvalues, $\mathbf{v}_j$ is the Principal Diffusion Direction (PDD), and $\mathbf{I}$ is the identity matrix. The algorithm iterates three steps, like \citet{sadd_ismrm, aranda2015}: Predict, Correct, and Generate, until convergence. Prediction selects the relevant DFs using LASSO to regulate the number to choose. Correction adjusts volume fraction, size, and orientation of the DTs. Taking advantage of DT expression and Taylor first order series approximation of the exponential, the optimisations are reduced to bounded Least Squares problems which are solved by a Projected Gauss-Seidel scheme. Generation controls the overestimation of fibers by adding to the basis the resulted DTs of combining two and three DFs for the new iteration.  

An extra refinement to the computed results, named LASADD--3C, splits each detected DF into three compartments \citep{sherbondy2010}: intracellular (IC), extracellular (EC) and CSF. The multi-tensor model is $s_i^{*} = \sum_{j=1}^n \alpha^{IC}_j \psi_{i,j} + \sum_{j=1}^n \alpha^{EC}_j \theta_{i,j} + \alpha^{CSF} \omega_{i}$ with $\sum_{j=1}^n \left( \alpha^{IC}_j + \alpha^{EC}_j \right)+\alpha^{CSF}=1$. The $\psi_{i,j}$ models the directional IC compartment diffusion for each fiber bundle
 using $\mathbf{T}^{IC}_j = {\chi_0}_j \mathbf{v}_j \mathbf{v}_j^T$. The EC compartment with hindered diffusion uses the representation \eqref{lasadd_tensor} for $\theta_{i,j}$. The isotropic diffusion $\omega_i$  uses $\mathbf{T}^{CSF}={\chi_3} \mathbf{I}$. This stage keeps fixed the PDDs and only adjust the $\alpha$'s and $\chi$'s of the three compartments.

The parameters of the models were estimated using the training dataset: the $b$ values using the equation by \citet
{stejskal_tanner65} and the ${S_0}$ values as the median of the gradient--free signals with equal echo time per voxel. The initial basis comprises 33 PDDs distributed in the unitary sphere. The bounds $\chi_{\{0,1\}} \in [1,39]\times10^{-4}$  and $\chi_{\{2,3\}} \in [1,9]\times10^{-4} \text{mm}^2/\text{s}$  and the LASSO regularisation parameter (equals $1.7$) were tuned by hand such that provides the minimum error. The best multi-tensorial model for both algorithms was used for each voxel to predict the corresponding unseen data.

\subsection{Signal models}
\subsubsection{Alipoor (Chalmers, Sweden)}
The DMRI signal is modeled as a fourth-order symmetric tensor as proposed by \citep{Ozarslan03}. Let $\mathbf{g}_i=[x_i~y_i~z_i]$ and $\mathbf{a}_i=[z_i^4~4y_iz_i^3~6y_i^2z_i^2~4y_i^3z_i~y_i^4~4x_iz_i^3$ $~12x_iy_iz_i^2~12x_iy_i^2z_i~4x_iy_i^3$ $~6x_i^2z_i^2~12x_i^2y_iz_i$ $~6x_i^2y_i^2$ $4x_i^3z_i~ 4x_i^3y_i~ x_i^4]^T$ be a gradient encoding direction and corresponding design vector, respectively. The diffusion signal is then described by
\begin{equation}
S(\mathbf{g}_i)=S_{0}\exp(\frac{-TE}{T_2})\exp(-b\mathbf{t}^{T}\mathbf{a}_i)  \label{e1}
\end{equation}
where $S(\mathbf{g}_i)$ is the measured signal when the diffusion sensitizing gradient is applied in the direction $\mathbf{g}_i$, $S_{0}$ is the observed signal in the absence of such a gradient, $b$ is the diffusion weighting factor, and $\mathbf{t}\in\mathbb{R}^{15}$ contains the distinct entries of a fourth-order symmetric tensor. Note that $d(\mathbf{g}_i$,$\mathbf{t})=d(\mathbf{g}_i)$ is used for simplification. Given measurements in $N>15$ different directions, the  least squares (LS) estimate of the diffusion tensor is $\mathbf{\hat{t}}=(\mathbf{G}^{T}\mathbf{G})^{-1}\mathbf{G}^{T}\mathbf{y}$
where $\mathbf{G}$ is an $N\times 15$ matrix defined $\mathbf{G}=[\mathbf{a}_{1}~\mathbf{a}_{2}\cdots~\mathbf{a}_{N}]^{T}$ and $y_i=-b^{-1}\ln(S(\mathbf{g}_i)/S_{0})$. We use the weighted LS tensor estimation method in \citep{EMBC13} to mitigate the influence of outliers. \\
To estimate the diffusion signal for a given acquisition protocol with $TE=TE_{x}$, $b=b_x$ and $\delta=\delta_x$, the two non-diffusion weighted measurements with the closest $TE$s to $TE_x$ (among measurements with $\delta=\delta_x$) are used to estimate $T_2$ and $S_0$ for each voxel. Then, data from the closet shell to $b_x$ (among shells with $\delta=\delta_x$) are used to estimate the tensor describing the underlying structure.

\subsubsection{Sakaie-Tatsuoka-Ghosh (Cleveland, USA): An Empirical Approach} 
As the extent of q-space in the dataset is unusually comprehensive, we chose a simple, generic approach to gain intuition. Visual inspection suggested use of a restricted and hindered component each with angular variation:
\begin{equation}
S_i = A_{TE_i}(fR_i + (1 - f)exp(-b_iD_i))
\end{equation}
where $S_i$ is the predicted signal for signal acquired with $TE_i$, $b_i$. $A_{TE_i}$ is the median signal at a given TE with no diffusion weighting. Fit parameters are $f$, the volume fraction of $R_i$, the restricted component, and $D_i$, the diffusivity. $R_i$ and $D_i$ are modeled as spherical harmonics with real, antipodal symmetry \citep{alexander2002detection} with maximum degree 4. The model has 31 fit parameters for each voxel. Data were fit using using a nonlinear least squares algorithm (lsqcurvefit, MATLAB). Prior to the fit, data points with nonzero bvalue that had signal higher than the the median of the b=0 signal plus 1.4826 times the median absolute deviation were excluded. Shells with normalized median signal smaller than that of shells with lower bvalues were also excluded. Normalization was performed by dividing by the median of the b=0 signal with the same TE.

\subsubsection{Fick (INRIA, France): A Spatio-Temporal Functional Basis to Represent the Diffusion MRI Signal}
We use our recently proposed spatio-temporal (3D+t) functional basis \citep{fick2015unifying} to simultaneously represent the diffusion MRI signal over three-dimensional wave vector \textbf{q} and diffusion time $\tau$. Based on Callaghan's theoretical model of spatio-temporal diffusion in pores \citep{callaghan1995pulsed}, our basis represents the 3D+t diffusion signal attenuation $E(\textbf{q},\tau)$ as a product of a spatial and temporal functional basis as 
\begin{equation}\label{eq:basisformulation}
 E(\textbf{q},\tau)=\sum_{N=0}^{N_{\textrm{max}}}\sum_{\{jlm\}}\sum_{o=0}^{O_{\textrm{max}}}c_{\{jlmo\}}\,S_{jlm}(\textbf{q},u_s)T_{o}(\tau, u_t)
\end{equation}
where $T_{o}$ is our temporal basis with basis order $o$ and $S_{jlm}$ is the spatial isotropic MAP-MRI basis \citep{ozarslan2013mean} with radial and angular basis orders $j$, $l$ and $m$. Here $N_{\textrm{max}}$ and $O_{\textrm{max}}$ are the maximum spatial and temporal order of the bases, which can be chosen independently. We formulate the bases themselves as
\begin{equation}\label{eq:basis}
\begin{aligned}
 S_{jlm}(\textbf{q},u_s)&=\sqrt{4\pi}i^{-l}(2\pi^2u_s^2q^2)^{l/2}e^{-2\pi^2u_s^2q^2} L_{j-1}^{l+1/2}(4\pi^2u_s^2q^2)Y_l^m(\textbf{u})\\
 T_o(\tau, u_t)&=\exp(-u_t\tau/2)L_o(u_t\tau)
\end{aligned}
\end{equation}
with $u_s$ and $u_t$ the spatial and temporal scaling functions, $Y_l^m$ the spherical harmonics and $L_o$ a Laguerre polynomial. We calculate the spatial scaling $u_s$ by fitting an isotropic tensor to the TE-normalized signal attenuation $E(\textbf{q},\cdot)$ for all $\textbf{q}$. Similarly, we compute $u_t$ by fitting an exponential $e^{-u_t\tau/2}$ to $E(\cdot,\tau)$ for all $\tau$. We fit our basis using Laplacian-regularized least squares in the following steps: We first denote $\Xi_i(\textbf{q},\tau,u_s,u_t)=S_{jlm(i)}(\textbf{q},u_s)T_{o(i)}(\tau,u_t)$ with $i\in\{1\ldots N_{\textrm{coef}}\}$ with $N_{\textrm{coef}}$ the number of fitted coefficients.
We then construct a design matrix $\textbf{Q}\in \mathbb{R}^{N_{\mathrm{data}}\times N_{\mathrm{coef}}}$ with $\textbf{Q}_{ik}=S_{N_i}(\textbf{A},\textbf{q}_k)T_{o_i}(\tau_k,u_t)$. The signal is then fitted as $\textbf{c}=\textrm{argmin}_{\textbf{c}}\|\textbf{y}-\textbf{Qc}\|^2+\lambda\,U(\textbf{c})$ with $\textbf{y}$ the measured signal, $\textbf{c}$ the fitted coefficients and $\lambda$ the weight for our analytic Laplacian regularization $U(\textbf{c})$.  We used generalized cross-validation \citep{craven1978smoothing} to find the optimal regularization weighting $\lambda$ in every voxel. In our submitted results, we used a spatial order of $8$ and a temporal order of $4$, resulting in $475$ fitted coefficients.\\

\subsubsection{Rivera (CIMAT, Mexico): Baseline Method: Robust Regression} 

We regard this very simplistic model as a baseline for other model-based methods. It assumes as little information as possible from the diffusion signal. 
The vector of independent variables is $x_i = [g_i, |G|_i, \Delta_i, \delta_i, TE_i, b_i]$, containing the gradient strength $g$, the echo time $TE$ and the b-value $b$. Given signal $s_i$, we then estimate the parameters of the linear regression model:
\begin{equation}
\label{eq:one}
	s= X \theta + \epsilon
\end{equation}
where $\theta \in R^{23}$ is the unknown vector of coefficients, $\epsilon$ is the residual error and  
$$
	X = \left[x, x|^2, \Delta \,\delta, \Delta \, TE , \Delta \, b,  \delta \, TE, \delta \, b ,TE \, b,  1 \right]
$$
is the \emph{matrix design} ($x|^2$ is obtained from squaring each element of the matrix x). To account for outliers we estimate $\theta$ with a weighted (robust) least squares approach using the Lasso Regularization: 
\begin{equation}
\label{eq:two}
 	\theta^{t+1}  =  {\underset{\theta}{\operatorname{argmin}}\;}
		  \| W^t(X \theta -y) \|_2^2 + \lambda \| \theta \|_1
\end{equation}
where $W^{0}$ is the identity matrix and each subsequent $W$ computed via:
\begin{equation}
\label{eq:three}
	W^{t+1} = \operatorname{diag}( v_i^{t+1} w_i^{t+1} )
\end{equation}
with outlier weighting in $\omega_i^{t+1} = {\kappa^2 } / ({\kappa^2 + (y_i -X\theta^{t+1}_i)^2}) $
though $\kappa$, an arbitrary parameter that controls the outlier sensitivity. The protocol weight
\begin{equation}
	v^{t+1}_i = \operatorname{mean}_{j \in \Omega_i} \{w^{t+1}_j\}  \;\mbox{and} \; \Omega_i = \{j: TE_j=TE_i,
|G_j|=|G_i| \}
\end{equation}
computes a confidence factor for the complete protocol. 

The equations \eqref{eq:two} and \eqref{eq:three} are iterated three times. The final estimated signal is computed using \eqref{eq:one}, using the protocol of the unseen signal.

\section*{References}

\bibliography{mybibfile}

 \begin{table}[htb]
\begin{minipage}[b]{1.0\linewidth}
  \centering
  \centerline{\includegraphics[width=12cm]{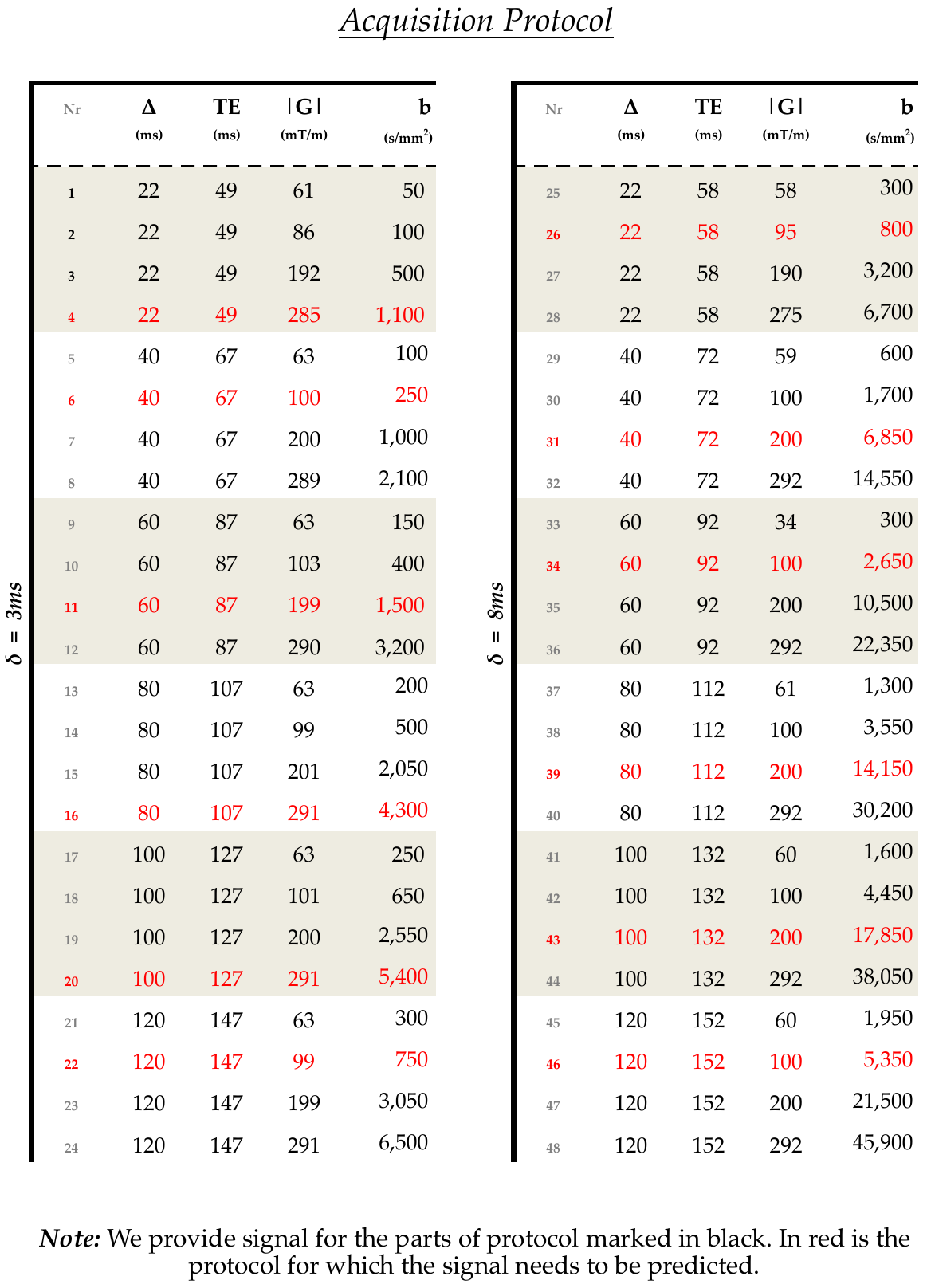}}
\end{minipage}
\caption[]{The scanning protocol used, acquired in $\sim$8 hours over two non-stop sessions. The protocol has 48 shells, each with 45 unique gradient directions (`blip-up-blip-down').}
\label{fig:protocol}
\end{table}

\begin{table}[tbp]
\centering
\small
\begin{tabular}{lccccccc} & 
\begin{tabular}[c]{@{}c@{}}Type of \\ model
\end{tabular} & 
\begin{tabular}[c]{@{}c@{}}Nb of free\\ param. \\ (genu/fornix)
\end{tabular} & 
\begin{tabular}[c]{@{}c@{}}Models\\ effect of\\ $\delta$ and $\Delta$
\end{tabular} & 
\begin{tabular}[c]{@{}c@{}}Noise \\ assumption
\end{tabular} & 
\begin{tabular}[c]{@{}c@{}}Optimization \\ algorithm
\end{tabular} & 
\begin{tabular}[c]{@{}c@{}}Outliers \\ strategy
\end{tabular} & 
\begin{tabular}[c]{@{}c@{}}Special \\ signal\\ prediction \\ strategy
\end{tabular} 
\\ \hline \hline
\begin{tabular}[c]{@{}l@{}}
R--Manzanares
\end{tabular} 
& Tissue & N/A & Yes & Gaussian  & 
\begin{tabular}[c]{@{}c@{}}weighted-LS\\  bootstrapping
\end{tabular}
& Yes & CV
\\ \hline
Nilsson & Tissue & $\textless$ 12/12 & Yes & Gaussian & 
\begin{tabular}[c]{@{}c@{}}LM
\end{tabular} 
& Yes & 
\begin{tabular}[c]{@{}c@{}} CV
\end{tabular}
\\ \hline
Scherrer & Tissue & 10/16 & No & Gaussian & Bobyqa & Yes & No  
\\ \hline
Ferizi\_1 & Tissue & $\textless$ 12/12 & Yes  & approx.-Rician  & 
\begin{tabular}[c]{@{}c@{}}LM
\end{tabular} & No & No  
\\ \hline
Ferizi\_2 & Tissue & $\textless$ 10/10 & Yes  & approx.-Rician  & 
\begin{tabular}[c]{@{}c@{}}LM
\end{tabular} & No & No  
\\ \hline
Alipoor     & Signal & 17/17     & No    & Gaussian       & 
\begin{tabular}[c]{@{}c@{}}weighted-LS
\end{tabular}  & Yes & No
\\ \hline
\begin{tabular}[c]{@{}l@{}}
Sakaie
\end{tabular}
& Signal & N/A & No & Gaussian & 
\begin{tabular}[c]{@{}c@{}}nonlinear-LS
\end{tabular} & Yes & No  
\\ \hline
Rokem & Tissue & $\sim$20  & No & 
\begin{tabular}[c]{@{}c@{}}Gaussian\\ + Noise floor
\end{tabular} & Elastic net  & No  & 
\begin{tabular}[c]{@{}c@{}}CV
\end{tabular}
\\ \hline
Eufracio & Tissue & 7/7    & No    & Gaussian & 
\begin{tabular}[c]{@{}c@{}}bounded-LS\\ Lasso, Ridge
\end{tabular}
& No & No
\\ \hline
Loya-Olivas\_1 & Tissue  & 11    & No    & Gaussian & 
\begin{tabular}[c]{@{}c@{}}bounded-LS\\ \& Lasso
\end{tabular}                  
& No  & No
\\ \hline
Loya-Olivas\_2 & Tissue  & 5    & No    & Gaussian & 
\begin{tabular}[c]{@{}c@{}}bounded-LS
\end{tabular}                  
& No  & No
\\ \hline
Poot & Signal  & 103    & No    & Rician & 
\begin{tabular}[c]{@{}c@{}} LM-like
\end{tabular}                  
& No  & No
\\ \hline
Fick & Signal & 475 & Yes   & Gaussian & 
\begin{tabular}[c]{@{}c@{}}Laplacian-\\ reg-LS
\end{tabular} 
& No  & 
\begin{tabular}[c]{@{}c@{}}partial-CV
\end{tabular} 
\\ \hline
Rivera & Signal & N/A & Yes   & Gaussian & 
\begin{tabular}[c]{@{}c@{}}Weighted Lasso
\end{tabular} 
& Yes  & 
\begin{tabular}[c]{@{}c@{}}CV
\end{tabular}

\end{tabular}
\normalsize
\caption{\label{table:modelsummary}Summary of the various diffusion models evaluated.
Tissue models are models that include an explicit description of the underlying tissue microstructure with a multi-compartment approach.
In contrast, signal models focus on describing the DW signal attenuation without explicitly describing the underlying tissue
and rather correspond to a ``signal processing'' approach.  (Abbreviations: LS=Least Squares, LM=Levenberg-Marquardt, CV=cross-validation, reg=regularized.)}

\end{table}
 -------------------------------------------------------------------------



\end{document}